**Effect of Block Asymmetry on the Crystallization of Double Crystalline Diblock Copolymers**

Chitrita Kundu and Ashok Kumar Dasmahapatra[*]

Department of Chemical Engineering, Indian Institute of Technology Guwahati, Guwahati –

781039, Assam, India

[*] Corresponding author: Phone: +91-361-258-2273; Fax: +91-361-258-2291; Email address: akdm@iitg.ernet.in

**ABSTRACT**

Monte Carlo simulation on the crystallization of double crystalline diblock copolymer unravels an intrinsic relationship between block asymmetry and crystallization behaviour. We model crystalline A-B diblock copolymer, wherein the melting temperature of A-block is higher than that of the B-block. We explore the composition dependent crystallization behaviour by varying the relative block length with weak and strong segregation strength between the blocks. In weak segregation limit, we observe that with increasing the composition of B-block, its crystallization temperature increases accompanying with higher crystallinity. In contrast, A-block crystallizes at a relatively low temperature along with the formation of thicker and larger crystallites with the increase in B-block composition. We attribute this non-intuitive crystallization trend to the dilution effect imposed by B-block. When the composition of the B-block is high enough, it acts like a "solvent" during the crystallization of A-block. A-block segments are more mobile and hence less facile to crystallize, resulting depression in crystallization temperature with the formation of thicker crystals. At strong segregation limit, crystallization and morphological development are governed by the confinement effect, rather than block asymmetry. Isothermal crystallization reveals that the crystallization follows a homogeneous nucleation mechanism with the formation of two dimensional crystals. Two-step, compared to one-step isothermal crystallization leads to the formation of thicker crystals of A-block due to the dilution effect of the B-block.

## 1. INTRODUCTION

The potential usage of block copolymers in modern technology makes scientists to be more fascinating about these materials as their property can be tailored by tuning molecular architecture.[1] Diblock copolymer consists of two thermodynamically incompatible blocks covalently bonded together which further self-assembled into various nanostructures.[1] Degree of polymerization ($N$) and Flory-Huggins interaction parameter ($\chi$) plays a critical role in determining the extent of separation. A critical value of $\chi N$ is considered to be 10.5, below and above of this value is termed as weak and strong segregation limit respectively.[2] A large variety of thermodynamically stable phases including lamellar structure, hexagonally packed cylinder, and body centred cubic phases are observed over a wide range of composition in diblock copolymers.[3]

The crystallization of diblock copolymer with one crystallizable block has been extensively studied in last few decades.[4-20] However, the complexity associated with the crystallization of double crystalline diblock copolymer is poorly understood. In crystalline-crystalline diblock copolymer, the final crystal morphology is driven by the competition between two blocks towards crystallization along with microphase separation. Diblock copolymer usually follows the sequential crystallization mechanism, where melting point difference between two blocks is relatively large.[21-29] During crystallization, high melting block crystallizes first creating spatial confinement for the crystallization of the low melting block resulting less crystallinity in the second block.[21-24, 26-34] However, depending on the nature of the blocks in diblock copolymer, they may crystallize simultaneously (viz., coincident crystallization).[28, 32, 35-38]

Primarily, the interplay between crystallization and microphase separation plays a crucial role in dictating the final crystal morphology of the semi crystalline materials.

Moreover, crystallization temperature ($T_c$), degree of immiscibility (viz., segregation strength, value of χ), cooling pathways (non-isothermal or isothermal) and block composition are the other pivotal factors which influence the crystallization of diblock copolymer to a large extent. By changing the block length ratio, one can easily manipulate the sequence of crystallization and crystal orientation influencing the final crystal morphology. For example, in linear poly(ethylene)-*b*-hydrogenated poly(norbornene) (LPE-*b*-hPN) diblock copolymer, where the melting points of two blocks are very close to each other (~150°C), the crystallization behaviour is largely tuned by the relative block composition. In hPN rich diblock copolymer, hPN crystallizes first whereas in LPE rich diblock copolymer, LPE crystallizes first followed by the hPN block.[33] Small angle X-ray scattering (SAXS) and wide angle X-ray scattering (WAXS) studies showed that the crystal orientation is equally influenced by the block length ratio. The first crystallizable block creates a structural template in which the crystallization of the second block is confined with perpendicular orientation.[34] Similar crystallization behaviour has been observed in case of poly(ethylene glycol)-*b*-poly(ε-caprolactone) (PEG-*b*-PCL) diblock copolymer, where block with higher composition crystallizes first.[39]

Several scenarios have been observed in the crystallization of asymmetric diblock copolymers. For example, PCL-*b*-PE diblock copolymer, when crystallized isothermally, PE block crystallizes first resulting lamellar morphology which confined crystallization of the PCL block.[23, 24] However, block composition strongly dictates the crystallization behaviour. When the composition of PE block is less than 56%, PCL block partially disrupts the lamellar morphology of already crystallized PE block, as the crystalline layers of PE blocks offer very soft confinement during crystallization of the PCL block. When the composition of PE block is greater than 76%, the lamellar morphology of crystalline PE block remains unaltered as it acts as a hard confinement during crystallization of the PCL block.[31] Isothermal crystallization of poly(*p*-dioxanone)-*b*-PCL (PPDX-*b*-PCL) diblock copolymer reveals that when the composition of the PCL block in diblock copolymer is high enough ($\sim 60 - 77$ %), the crystallization rate of PCL block (in terms of crystallization half time) is increased, as the previously crystallized PPDX block acts as a nucleating agent to accelerate the crystallization of PCL block. However, when the composition of PCL block is $\leq 50\%$, the previously crystallized PPDX block imposes topological restriction (viz., confinement) in slowing down crystallization of the PCL block. Although the hydrolytic degradation of PPDX is reduced by PCL block, but the increased resistance to hydrolysis is a complex function of block composition.[37] Degree of asymmetry in the block composition influences crystallization and melting temperatures of the constituent blocks. For example, poly (L-lactic acid) (PLLA) block in PLLA-*b*-PCL diblock copolymer exhibits higher crystallization and melting temperature with increasing PLLA content. On the other hand, at lower compositions of PLLA, the reverse phenomena are observed with the reduced crystallization rate, which has been attributed to the fact that the PCL block (major component) act as a diluent, and causes the depression of crystallization and melting temperatures of PLLA block.[27, 29] Kinetic analysis based on Avrami equation indicates the development of a 3-dimensional sphurelitic superstructure with Avrami index $2.5 - 3$ for most of the compositions, whereas the diblock copolymer containing 10 wt.% PLLA forms axialites (viz., non-spherical and irregular superstructures) with Avrami index $1.0$.[27] Similar diluent effects on the PLLA blocks have been observed in PLLA-*b*-PEO diblock copolymer with increasing the composition of poly(ethylene oxide) (PEO) block.[40, 41]

In a recent work we have demonstrated the effect of segregation strength on the development of crystallinity in symmetric diblock copolymers.[42] We have observed that with increasing value of segregation strength, the degree of crystallinity gradually decreases producing smaller and thinner crystals. In the present study, we report Monte Carlo simulation results on the effect of block composition on the crystallization and morphological development of a series of diblock copolymers. We present our results based on two different levels of segregation. At weak segregation, we observe a non-monotonic trend in the lamellar thickness of A-block with the increasing composition of B-block, which is attributed to the dilution effect of the B-block.

We organize our paper as follows. We describe the model and simulation technique in section 2 followed by results and discussion in section 3 and conclusion in section 4.

## 2. MODELLING AND SIMULATION TECHNIQUE

Monte Carlo simulation has been evolved as a powerful tool to investigate the phase behaviour of polymeric systems during the last few decades. Several Monte Carlo techniques such as Pruned Enriched Rosenbluth Method (PERM),[43] random end-switch-configuration biased Monte Carlo (RES-CBMC),[44] and wormhole algorithm[45] are successfully applied to simulate a single polymer chain with some limitations for dense systems. Lattice simulations of dense systems have successfully been implemented by Pakula et al.[46] with the cooperative motion algorithm (COM)[47] and Hu et al.[48] with the single site bond fluctuation algorithm.[49-51] In cooperative motion algorithm (CMA), the occupation density of lattice polymer is 100%, however, the morphological evolution with time cannot be explicitly mapped with Monte Carlo steps; whereas by using bond fluctuation model, Hu et al. produced the experimentally observed nontrivial trend of copolymer crystallization.[52] Moreover, Hu et al. have studied the effect of sequence distribution of comonomer on crystallization of different statistical copolymers,[52] Single site bond fluctuation algorithm have been successfully employed to investigate the effect of sticky additives in polymer crystallization[53] and the crystallization of double crystalline diblock copolymer where the crystallization of one block accelerates the crystallization of other block.[54]  In the present work, we employ dynamic Monte Carlo simulation to study the crystallization of double crystalline diblock copolymer from a homogeneous melt.

Initially, we place 480 polymer chains in a cubic lattice having 32×32×32 lattice units, with a lattice occupation density 0.9375 that represents a polymer bulk system.  The degree of polymerization ($N$) of a polymer chain is equal to 64 including $N_A$ numbers of A-block and $N_B$ numbers of B-block units.  We generate an initial structure by placing block units one by one along the lattice grid.  We express the composition of A- and B-block by $x_A$ ($N_A/N$) and $x_B$ ($N_B/N$) respectively.  The length of each block (viz., A-block and B-block) in a polymer chain can be manipulated by $x_B$.  We multiply $x_B$ with the degree of polymerization ($N$) to get the composition of B-block.  For example, in a simulation box where $x_B$ is 0.25, per chain number of units of B-block is 16 (viz., 0.25×64) and remaining 48 units (viz., 0.75×64) are of A-block.  Then, we apply a set of microrelaxation algorithms including bond fluctuation, end bond rotation and slithering diffusion to move the chain units along the lattice sites.[42, 52, 53, 55-57]  To give further details, we start with finding a vacant site randomly from the available vacant sites and then we select a neighbouring occupied site (occupied by either A- or B-type unit) to initiate a microrelaxation move.  According to the position of the block units (A- or B-type) along the chain, we choose the appropriate algorithm.  For terminal units, the end bond rotation and slithering diffusion are implemented with equal probability; and for non-terminal one, single site bond fluctuation is implemented.  The coordination number of our lattice model is 26, which allows a bond to possess a value equal  to 1 (along the axis), $\sqrt{2}$ (along the face diagonal) or $\sqrt{3}$ (along the body diagonal) units.[42, 52, 53]

The mutual immiscibility between two blocks in diblock copolymer is modelled by a repulsive interaction, $U_{AB}$. The crystallization driving force is modelled as an attractive interaction between neighbouring parallel and collinear bonds, which are represented by $U_p$ and $U_c$ respectively. The change in energy per Monte Carlo (MC) move is then modelled by:

$$\Delta E = -\left(\Delta N_p U_p + \Delta N_c U_c\right)_A - \left(\Delta N_p U_p + \Delta N_c U_c\right)_B + \Delta N_{AB} U_{AB}$$

Where, $\Delta N_p$ and $\Delta N_c$ represents the net change in the number of parallel and collinear bond respectively for the A- and B-block respectively, and $\Delta N_{AB}$ represents the change in the number of contacts between A and B units after each microrelaxation move.[42]

To model the difference in melting temperature between two blocks, we use $U_{pB} = \lambda_m U_{pA}$ and $U_{cB} = \lambda_m U_{cA}$ for the parallel and collinear bond respectively. Now, we set $\lambda_m = 0.75$ ($< 1$) to represent B-block as a low melting block and hence, less crystallisable compared to A-block. We also set $U_p = U_c$ for both the blocks to represent coarse grained interaction in our system[53]. We calculate $U_{AB}$ as $\lambda U_p$, where $\lambda$ represents the segregation strength (viz., degree of immiscibility) between two blocks and $\lambda > 0$. Higher the value of $\lambda$, higher is the extent of immiscibility between the blocks. We set $\lambda = 1$ and 4 to represent weak and strong segregation limit[42]. According to the Flory-Huggins interaction parameter, the value of $(\chi N)$ determines the extent of segregation. The value of $(\chi N)$ can be mapped with $(q-2) \times U_{AB} \times N$ of our system where $q$ is the coordination number and $N$ is the degree of polymerization[42]. All the energies are normalized by $k_{\beta T}$ and $U_p \sim 1/T$ where, $k_B$ is the Boltzmann constant. Now the change in energy per MC move is modified as follows:

$$\Delta E = \left[ -\left( \Delta N_p + \Delta N_c \right)_A - \lambda_m \left( \Delta N_p + \Delta N_c \right)_B + \lambda \Delta N_{AB} \right] U_p$$

We have implemented the Metropolis sampling scheme with periodic boundary condition to select a new state with a probability equal to $\exp(-\Delta E)$. We accept new conformation if $\exp(-\Delta E) \le r$, where $r$ is the random number (0, 1) generated by the random number generator, MT19937.[58]  We equilibrate our sample system for 5000 Monte Carlo steps ($MCS$), and the thermodynamics and structural parameters are calculated averages over subsequent 5000 $MCS$. One MCS is defined as 480 × 64 MC moves, viz., on the average one attempted MC move for each unit, A-type and B-type present in the simulation box.

To follow transition from a disordered melt to an ordered crystalline phase, we calculate fractional crystallinity for A-block ($X_A$), B-block ($X_B$) and overall ($X_c$) as a function of $U_p$. We measure crystallinity as the ratio of number crystalline bonds to the total number of bonds in the system. If a bond is surrounded by more than 5 nearest non-bonded parallel bonds, it is considered as a crystalline bond.[42, 52, 53]  To calculate $X_A$ and $X_B$, we consider A-block and B-block units present in the system, respectively. To locate the transition point from melt to crystal, we calculate specific heat ($C_v$) as equilibrium specific heat from the total energy fluctuations in the simulation box. [42, 52, 53]  To locate the microphase separation point, we calculate $C_v$ of A-B pair considering the de-mixing energy between A and B-block. To monitor the relative mobility of A- and B-block during crystallization, we calculate mean square displacement of the centre of mass ($d_{cm}^2$) of each block, averaged over all the chains in the system. We also monitor the change in the average crystallite size $\langle S \rangle$ and lamellar thickness $\langle l \rangle$ during crystallization. A crystallite is a small microscopic aggregate having crystalline bonds in the same orientation. The crystallite size is measured as the total number of crystalline bonds present in it. The lamellar thickness is the average number of block units (A- or B-type) towards the direction of crystal thickness in a given crystallite, and their average is calculated over all the crystallites present in the system.

## 3. RESULTS AND DISCUSSIONS

We begin by exploring the effect of block composition (viz., $x_B$ = 0.125, 0.25, 0.375, 0.5, 0.625, 0.75 and 0.875) on crystallization and morphological development of double crystalline diblock copolymer. We limit our discussion with two levels of segregation strengths: $\lambda$ = 1 for weak and $\lambda$ = 4 for strong segregation. Subsequently, we discuss transition kinetics over a wide range of block composition by isothermal crystallization with both the segregation levels.

### 3.1. Monitoring Phase Transition

To simulate crystallization of diblock copolymer, we generate an equilibrated high temperature melt at $U_p$ = 0 (T = ∞, athermal state), where both the blocks are homogeneously mixed. Figure 1 displays the snapshots of evenly dispersed melt composed of A- and B-type units at $x_B$ = 0.25 and 0.75 respectively (blue lines represent A-block and orange lines represent B-block segments). The snapshots clearly display an isotropic orientation of polymer chains in both the compositions at $\lambda = 1$. For the snapshots of homogeneous melt of the other block compositions, see Figure S1, supplementary information.[59] We cool the sample system from $U_p = 0$ to $U_p = 0.6$ with a step size of 0.02 by implementing the non-isothermal cooling process which results in parallel alignment of polymer chains indicating ordered crystalline regions. We monitor the transition of diblock copolymer from a homogeneous melt to a crystalline state by following the change in specific heat ($C_v$)[42, 53] calculated from energy fluctuations as a function of $U_p$. During crystallization, $C_v$ shows a peak at a certain value of $U_p$ considered as a transition point from a homogeneous disordered melt to an ordered crystalline state.[60] We plot $C_v$ vs. $U_p$ over a wide range of block compositions (viz., $x_B = 0.125$ to $x_B = 0.875$) in Figure 2 at two different levels of segregation strength (viz., $\lambda = 1$ and 4). Since the crystallization driving force of the A-block is higher than the B-block, they follow a sequential crystallization mechanism (viz., they crystallize separately), which is reflected by the presence of two different transition peaks at two different values of $U_p$. This observation is in accord with the experimental results on PE-*b*-PEO diblock copolymer which gives sequential crystallization at two different temperatures (viz., 95.4 °C for PE block and 12.9 °C for PEO block).[21] We summarize the change in crystallization temperatures ($U_p^*$) of two blocks in weak segregation limit in the inset of Figure 2a. We observe that the crystallization temperatures of A-block ($U_{pA}^*$) remain same for all block compositions except at $x_B = 0.875$, where it shows a relatively higher value of $U_p$ ($U_p^* = 0.3$). It appears that there is a small depression in crystallization temperature ($U_p \sim 1/T$), of A-block at high block composition (viz., $x_B = 0.875$). As the primary stage of crystallization is driven by the thermodynamic driving forces (viz., degree of cooling), the transition point of A-block does not vary with most of the block compositions, because the crystallization driving force ($\Delta T$) is similar for all the compositions investigated. However, at $x_B = 0.875$, a small depression in crystallization temperature of A-block is observed. The reason for this non-intuitive and unexpected trend in crystallization temperature is attributed to the dilution effect of the B-block. At higher $x_B$, B-block acts like a "solvent" for A-block. We observe that during the crystallization of A-block (viz., $U_p \sim 0.3$), the mean square displacement of the centre of mass of the A-block ($d_{cm}^2$ of A-block) increases at higher $x_B$ (Figure 3a) compared to that of the B-block. Higher mobility makes the chain segments less facile to crystallize at that temperature (viz., thermal driving force). On lowering the temperature further (viz., higher $U_p$), A-block crystallizes owing to the fact that it has now higher thermal driving force towards crystallization. On the other hand, we observe that the crystallization temperature of B-block increases (viz., decrease of $U_p$) with increasing block length (Figure 2a, inset), and the mean square displacement of the centre of mass of the B-block ($d_{cm}^2$ of B-block) decreases with composition (Figure 3a). The presence of a higher proportion of the B-block facilitates the formation of larger size domain with reduced chain mobility; and as a result, crystallization happens at a relatively higher temperature while the thermodynamic driving force for the crystallization is relatively less (viz., at a higher temperature). This observation is in accord with the experimental results of PLLA-*b*-PCL diblock copolymer where the crystallization temperature of PLLA block increases with the increased composition of PLLA-block.[27, 29] In the PEO-*b*-PCL diblock copolymer, the crystallization temperature of PCL block also increases with an increasing weight fraction of PCL block in the copolymer.[61] Similarly, in PLLA-*b*-PEG diblock copolymer, the crystallization temperature of PLLA block increases with increasing molecular weight of PLLA block.[40] We also estimate the transition points in terms of $U_p$ of both the blocks at $\lambda$

= 4 (higher segregation strength) in the Figure 2b. The overall trend of $C_v$ vs. $U_p$ appears to be identical to that of $\lambda = 1$. However, the transition points of both the blocks show a non-monotonic trend with composition (Figure 2b, inset), which is attributed to the effect of strong segregation, which causes hard confinement and creating large numbers of microdomains in the phase separated melt. Due to the strong confinement, chain mobility of both the blocks is highly restricted at strong segregation (viz., $\lambda = 4$), which is in accord with Figure 3.

### 3.2. Locating Microphase Separation

Usually in diblock copolymer, immiscibility between two blocks leads to the formation of microphase separated melt with various morphological patterns.[6-10, 13-16, 23, 24, 28, 29, 32, 38] Therefore, energy fluctuation based on A-B contacts would also exhibit a peak at the microphase separation point.[42] We compare microphase separation points ($U_p^\#$) with block compositions ($x_B$) in Figure 4 at two different segregation levels. At weak segregation (viz., $\lambda = 1$), the relative location of microphase separation point vary widely with block composition of B ($x_B$) but at strong segregation (viz., $\lambda = 4$), the microphase separation points are $\sim 0.01$, nearly independent of block composition of B ($x_B$). When the segregation strength of the system is high, composition imposes a marginal effect on microphase separation point, because segregation between two blocks plays the pivotal role to control the microphase separation. At weak segregation, the extent of microphase separation is dominated by the block length (viz., composition), whereas, at strong segregation, strength of segregation dominates over composition. As we can see from Figure 4, the value of $U_p^\#$ decreases from 0.052 for $x_B = 0.125$ to 0.016 for $x_B = 0.50$ and again 0.05 for $x_B = 0.875$.

From the above data, it appears that the microphase separation is retarded (viz., happens at a lower temperature) with the increase of degree of asymmetry in the diblock copolymer. This observation is in accordance with the melt behaviour of ethylene-*b*-ethylethylene (E-*b*-EE) diblock copolymer, wherein a composition-dependent microphase separation (measured in terms of order-disorder transition temperature, $T_{ODT}$) is observed. Diblock copolymers with 0.25, 0.49 and 0.75 weight fraction of ethylene exhibit $T_{ODT}$ at 255, 121 and 148 °C respectively.[6] Similarly, PLLA-*b*-PCL diblock copolymer exhibits $T_{ODT}$ at 175 and 220 °C for the sample having 37.4 and 46 wt% PCL block respectively.[62] The snapshots of microphase separated melt for $x_B = 0.25$ and 0.75 at two different segregation limits at $U_p = 0.1$ are shown in Figure 5, wherein the formations of phase segregated microdomains are clearly visible. The snapshots of microphase separated melt (at $U_p = 0.1$) for other compositions at $\lambda = 1$ and 4 are available in Figure S2 and S3 (Supplementary information),[59] respectively.

### 3.3. Evolution of Crystallinity in Non-isothermal Crystallization

We study the development of crystallinity during non-isothermal crystallization as a function of block composition at two different segregation levels. We present the change in overall crystallinity ($X_c$) at weak segregation as a function $U_p$ in Figure 6a. Overall crystallinity is calculated as the weighted average of the summation of crystallinity of A- and B-block (viz., $X_c = x_A \times X_A + x_B \times X_B$). With the increased value of $U_p$, the overall crystallinity gradually increases and at $U_p \sim 0.3$, it shows an abrupt change in its value; and finally it reaches to a plateau at $U_p \sim 0.5$ (cf., saturation crystallinity) where there is no significant changes in crystallinity. The change in crystallinity of A-block ($X_A$) and B-block ($X_B$) with $U_p$ (for $\lambda$ = 1 and 4) is presented in Figure S4 and S5 respectively.[59] The saturation crystallinity of A-block, B-block and overall at $U_p$ = 0.6 are plotted in Figure 6b as a function of block composition of B ($x_B$). At weak segregation, overall crystallinity does not vary too much with the change in block composition (Figure 6b). The values are ~ 0.70 for all the compositions investigated. This happens because of the driving force of the initial stage of crystallization is primarily dominated by the degree of cooling. The saturation crystallinity of A-block is almost independent of block composition, whereas the saturation crystallinity of B-block monotonically increases with the increasing composition of the B-block (viz., $x_B$). The values of saturation crystallinity of A-block at $\lambda$ = 1 are 0.72, 0.71, 0.71, 0.70, 0.70, 0.71 and 0.71 for $x_B$ = 0.125, 0.25, 0.375, 0.50, 0.625, 0.75 and 0.875 respectively. As we simulate each composition within the same degree of cooling, the crystallization driving force in each case is equal. Moreover, dilution effect of the B-block facilitates the formation of higher crystalline structure of A-block (see section 3.1) with increasing $x_B$. Therefore, the development of crystallinity of A-block over a wide range of composition appears to be independent of $x_B$. On the other hand, the saturation crystallinity of the B-block at $\lambda$ = 1 is 0.60, 0.63, 0.66, 0.66, 0.67, 0.68 and 0.69 for $x_B$ = 0.125, 0.25, 0.375, 0.50, 0.625, 0.75 and 0.875 respectively. As the A-block is more crystallisable than the B-block, the saturation crystallinity of B-block is always less than the A-block. Moreover, during crystallization, A-block creates some spatial confinements that suppress the crystallization of the B-block. With increased value of block composition of B, the saturation crystallinity of B-block increases due to the enhanced number of B-block units. This observation is in accord with the experimental results on the crystallization of PCL-$b$-PE[31] and PEO-$b$-PCL[61] diblock copolymers, wherein the crystallinity of PE and PCL block increases with increasing their content in PCL-*b*-PE and PEO-*b*-PCL diblock copolymer respectively. At strong segregation (viz., $\lambda = 4$), we observe that the overall crystallinity spreads over a wide range (0.3 to 0.6) with non-monotonic trend as a function of $x_B$ (Figure 7b). Similarly, the crystallinity of A-block and B-block also exhibit a non-monotonic trend with $x_B$ (Figure 7b). The values of saturation crystallinity of A-block at $\lambda = 4$ are 0.56, 0.60, 0.50, 0.36, 0.43, 0.46 and 0.25 for $x_B = $ 0.125, 0.25, 0.375, 0.50, 0.625, 0.75 and 0.875 respectively. The saturation crystallinity at $\lambda = 4$ is always less than the saturation crystallinity at $\lambda = 1$ for all the block compositions investigated. At $\lambda = 4$, the inter-block segregation strength is high enough to effectively confine the crystallization within the large number of microdomains created during microphase separation, and as a result crystallization is strongly inhibited. The values of saturation crystallinity of the B-block at $\lambda = 4$ are 0.15, 0.44, 0.31, 0.23, 0.36, 0.50 and 0.41 for $x_B = $ 0.125, 0.25, 0.375, 0.50, 0.625, 0.75 and 0.875 respectively. The snapshots for $x_B = $ 0.25 and 0.75 at the end of crystallization (viz., at $U_p = $ 0.6) are presented in Figure 8 and 9 at $\lambda = 1$ and 4 respectively. The blue and orange lines represent the crystalline segments of A- and B-block respectively, and yellow lines represent the non-crystalline segments of both the blocks. The snapshots clearly demonstrate that at strong segregation, the crystallinity decreases regardless of composition. The snapshots of the remaining compositions at $\lambda = 1$ and 4 are presented in Figure S6 and S7 (Supplementary information)[59] respectively.

### 3.4. Structural Analysis

We calculate average crystallites size and lamellar thickness separately for both the blocks as a function of $U_p$ for all the block compositions investigated. There is a wider distribution in crystallites size than lamellar thickness, and the magnitude of lamellar thickness is smaller compared to crystallites size, indicating the development of two dimensional crystals. Figure 10a represents the variation of average crystallite size (at $U_p$ = 0.6) as a function of $x_B$ for $\lambda$ = 1. At weak segregation (viz., $\lambda$ = 1), the crystallite size of A-block exhibits a non-monotonic trend, whereas the crystallite size of B-block monotonically increases with $x_B$ (Figure 10a). The reason behind this non-monotonic trend in A-block crystallites is attributed to the dilution effect of B-block (see section 3.1). When $x_B$ is low (viz., $x_A > x_B$), the formation of larger crystallites of A-block is naturally facilitated since A-block is more crystallizable than B-block. As $x_B$ increases, the relative stability of microdomains of B-block in the micropahse separated melt increases; and as a result, the formation of larger size crystallites of A-block is restricted. This observation is in line with SAXS results on PLLA-*b*-PCL diblock copolymer, which shows decreasing domain spacings of the PLLA block on increasing its content, increasing confinement for the crystallization of PCL block.[27] On further increasing of $x_B$ (viz., $x_B > 0.5$), the B-block now acts as the continuous phase with A-block as the disperse one. During crystallization of A-block ($U_p \sim 0.3$), B-block is in a molten state, and behaves like a "solvent", diminishing the topological restriction, which is reflected in the enhancement of mean square displacement of centre of mass (Figure 3). This enhanced mobility of A-block units favours the formation of larger size crystallites (only few crystallites could form, since $x_A < x_B$) as is shown in Figure 10a. The change in $\langle S_A \rangle$ and $\langle S_B \rangle$ vs. $U_p$ at different $x_B$ are presented in Figure S8 and S9 (supplementary information)[59] respectively.

To better understand the morphological evolution during crystallization, we also examine the average lamellar thickness of both the blocks as a function of compositions (Figure 10b). Figure S10 and S11 (supplementary information)[59] present the variation of $\langle l_A \rangle$ vs. $U_p$ and $\langle l_B \rangle$ vs. $U_p$ respectively. At weak segregation (viz., $\lambda = 1$), $\langle l_A \rangle$ does not vary appreciably with block compositions ($x_B$ = 0.125, 0.25, 0.375 and 0.5, Figure 10b). However, at higher values of $x_B$ ($x_B > 0.5$), the value of $\langle l_A \rangle$ significantly increases, and is attributed to the dilution effect of the B-block. When the composition of the B-block is less than 0.5, the crystallization of A-block is typically influenced by the chain entanglement and follows "melt crystallization", where the diffusion of chain segment is hindered by intra- and inter-chain entanglement, favouring folded chain crystals. However, when $x_B$ increases, A-block crystallizes within a matrix of the molten B-block, which act like a "solvent". The molten B-block imposes a marginal hindrance towards the diffusion of chain segments of A-block, and hence favouring extended chain crystals, which is a typical crystallization mechanism from a dilute solution. As a result, A-block crystallites thickens, giving rise to thicker crystals with larger crystallites size (as discussed above). Enhanced values of $\langle S \rangle$ and $\langle l \rangle$ well match with the magnitude of crystallinity at higher $x_B$ (Figure 6). The average crystallites size $\langle S \rangle$ and lamellar thickness $\langle l \rangle$ of both the blocks follows a non-monotonic trend with block compositions at strong segregation (Figure 11a and 11b).

### 3.5. Isothermal Crystallization

To follow the kinetic pathway of crystallization, we equilibrate our sample system at $U_p = 0$, quench from $U_p = 0$ to $U_p = 0.6$ and allow the system to anneal for $10^5$ Monte Carlo steps ($MCS$). Figure 12 represents the overall crystallinity ($X_c$) as a function of $MCS$ for all the block compositions investigated. Overall crystallinity is calculated by the summation of weighted average crystallinity of A- and B-blocks. The overall crystallinity shows a non-monotonic trend with compositions at both the levels of segregation; however, it establishes the dependence of transition kinetics on composition. The development of crystallinity with *MCS* in isothermal cooling of A-and B-block is presented Figure S12 and Figure S13 (Supplementary information)[59] at $\lambda$ = 1 and 4 respectively. During crystallization, as the driving force (viz., in terms of the temperature difference) is adequate for both the blocks, they compete for crystallization that leads to the coincident crystallization. This observation is in line with the isothermal crystallization of PPDX-*b*-PCL diblock copolymer, where crystallization kinetics of both the blocks is overlapped.[35, 36]

To get an insight on transition kinetics, nucleation type (viz., homogeneous or heterogeneous) and crystal geometry, we analyse the isothermal crystallization with the help of the Avrami equation.[63]

$$(1 - X) = \exp(-kt^n)$$

Where, $X$ is the overall crystallinity, $k$ and $n$ are constants. The value of Avrami index ($n$) indicates the type of crystal geometry. We estimate the Avrami index ($n$) based on the primary crystallization, which is considered as the development of crystallinity up to 20%.[64] At the weak segregation limit, the values of Avrami index of both the blocks are within the limit of 0.5 to 1.5 (Figure 13a) which indicates a first order transition kinetics with homogenous nucleation; whereas at strong segregation, due to confinement effect the value is decreased to the range of 0.5 to 1 (Figure 13b). This result is in line with the experimental observation on the crystallization of asymmetric double crystalline diblock copolymer, PLLA-*b*-PCL, which follows a homogeneous nucleation pathway with Avrami index close to 1.0.[27] Similar phenomena have also been observed in the crystallization of crystalline-amorphous diblock copolymers, where crystalline block is confined within the microdomains of amorphous block, follows a homogeneous nucleation pathway with the Avrami index ~

1.0. For example, crystallization of PEO block in poly(ethylene oxide)-*b*-poly(styrene) (PEO-*b*-PS) or PCL block in PCL-*b*-PS diblock copolymer, confined into large numbers of isolated microdomains of PS block (viz., spheres or cylinders) crystallizes via homogeneous nucleation.[16] PLLA block in PLLA-*b*-PS diblock copolymer[65] and PE block in styrene-*b*-ethylene-*b*-butane random terpolymer[66] also follow a similar mechanism with the Avrami index ~ 1.0.

In order to get an insight on the effect of quench depth on the development of crystalline structure, we carried out isothermal crystallization in two steps: we quench our sample system from $U_p = 0$ to $U_p = 0.3$ in the first step, and from $U_p = 0.3$ to $U_p = 0.6$ in the second step, with annealing for $10^5$ MCS in each step (viz., at $U_p = 0.3$ and $U_p = 0.6$). During the first step of quenching at $U_p = 0.3$ (temperature below the melting point of A-block but above the melting point of B-block) only A-block can crystallize and B-block is still in a molten state. During the second stage of quenching at $U_p = 0.6$ (temperature below the melting points of both the blocks), B-block crystallizes within the confined space created by A-block. Therefore, the mode of crystallization in two-step isothermal cooling is sequential crystallization, in contrast to the sequential crystallization in one-step cooling. Table 1 presents the crystallinity of A-block during one- and two-step isothermal crystallization of various block compositions at $\lambda = 1$ (viz., weak segregation). The crystallinity of A-block is close to 0.71 for $x_B$ up to 0.625 in two–step isothermal cooling, whereas, for $x_B = 0.75$ and 0.875, the crystallinity of A-block increases to 0.78. As the initial development of crystallinity is largely dominated by the degree of cooling, the crystallization of A-block is unaffected up to a certain block composition. However, at a very high value of $x_B$, dilution effect of B-block dominates the system, which reduces the topological restriction resulting increase in crystallinity of A-block. In the one-step isothermal cooling, the dilution effect is not prominent at higher composition. This observation is basically related to the change in mode of crystallization. In two–step isothermal crystallization, we observe a sequential crystallization similar to the non-isothermal crystallization. However, in one-step isothermal crystallization, we observe a coincident crystallization, where both the blocks compete for crystallization along with microphase separation. Table 2 compares the crystallinity of the B-block in two-step and one-step isothermal cooling at $\lambda = 1$. There is a monotonic increase in crystallinity of B-block with an increasing block length which is in line with the previous observation of non-isothermal crystallization. Figure 14 and 15 present snapshots from our simulation at $\lambda = 1$ and $x_B = 0.25$ and 0.75 respectively, for two-step and one-step isothermal cooling. The final crystalline structures for other compositions at $\lambda = 1$ are available in Figure S14[59] for isothermal two-step cooling and Figure S15[59] for isothermal one step cooling. However, at strong segregation (viz., $\lambda = 4$), the development of crystallinity of both the blocks is very less which ranges from 3 to 10% in one-step and 3 to 30% in two–step isothermal cooling. Therefore, it appears that the effect of block asymmetry on the development of crystallinity at strong segregation is negligible as most of the sample systems produce nearly amorphous structure in isothermal cooling (Table S1 and Table S2).[59]

## 4. CONCLUSIONS

We report the effect of block asymmetry on crystallization of double crystalline diblock copolymer by dynamic Monte Carlo simulation at two different levels of segregations. At weak segregation (viz., $\lambda = 1$), during non-isothermal cooling, diblock copolymer follows sequential crystallization mechanism regardless of composition. We observe a small depression in crystallization temperature of A-block at $x_B = 0.875$, which is attributed to the dilution effect of the B-block. We also observe a significant increase in chain mobility (in terms of mean square displacement of the centre of mass) of A-block at $U_p = 0.3$ compared to that of B-block at higher $x_B$. The crystallization temperature of the B-block monotonically increases with increasing block composition. At strong segregation, the transition points of both the blocks exhibit non-monotonic trend with block composition due to confinement effects dominated by large numbers of microdomains in microphase separated melt. The crystallinity of A-block remains identical over a wide range of block composition whereas the crystallinity of B-block increases with increasing block length of B at weak segregation. However the crystallinity of both the blocks produce non-monotonic trend with $x_B$ at strong segregation. The lamellar thickness of A-block significantly increases for $x_B = 0.75$ and $0.875$ at $\lambda = 1$ which is attributed to the dilution effect imposed by B-block. When the composition of A-block is very less in the system, B-block behaves as a "diluent", which reduces topological restriction favouring crystal thickening. Isothermal crystallization confirms the dependence of the compositions on transition kinetics at both the levels of segregations. The Avrami indexes demonstrate the presence of homogeneous nucleation with the formation of two-dimensional crystals. The dilution effect is more prominent in two-stage isothermal crystallization compared to one-stage crystallization due to the change in mode of crystallization. Two-stage isothermal cooling follows sequential crystallization whereas one-stage cooling follows coincident crystallization. Thus, manipulating block asymmetry with a proper choice of crystallization pathway, desired morphological pattern can be achieved. Present findings on the composition-dependent morphological development would enable in gaining insight in tailoring supramolecular properties of crystalline diblock copolymer for targeted applications.

**ACKNOWLEDGMENT**

We acknowledge the financial support from the Department of Science and Technology (DST), Government of India (sanction letter no. SR/S3/CE/0069/2010).

**Table captions**

**Table 1. Comparison in fractional crystallinity ($X_A$) of A-block with block compositions of B ($x_B$), during two-stage and one-stage isothermal crystallization at $\lambda$ =1.**

**Table 2. Comparison in fractional crystallinity ($X_B$) of B-block with block compositions of B ($x_B$) during two-stage and one-stage isothermal crystallization at $\lambda$ =1.**

**Figure captions**

**Figure 1. Snapshots of the simulation at $U_p$ = 0 representing homogeneous melt of diblock copolymer for (a) $x_B$ = 0.25 and (b) $x_B$ = 0.75. Blue and orange line represents segments of A- and B-block respectively.**

**Figure 2. Change in specific heat ($C_v$) with $U_p$ of different block composition ($x_B$) at (a) $\lambda$ = 1 and (b) $\lambda$ = 4; the inset shows the change in transition point ($U_p^*$) for A- and B-block with $x_B$. The lines joining the points are meant only as a guide to the eye.**

**Figure 3. Change in mean square displacement of centre of mass ($d_{cm}^2$) of A- and B-block with $x_B$ at (a) $U_p$ = 0.3 and (b) $U_p$ = 0.6. The lines joining the points are meant only as a guide to the eye.**

**Figure 4. Change in microphase Separation point ($U_p^{\#}$) with block composition ($x_B$) at two different segregation levels.**

**Figure 5. Snapshots of microphase separated melt at $U_p$ = 0.1 for (a) $x_B$ = 0.25 at $\lambda$ = 1, (b) $x_B$ = 0.25 at $\lambda$ = 4, (c) $x_B$ = 0.75 at $\lambda$ = 1 and (d) $x_B$ = 0.75 at $\lambda$ = 4. Blue and orange line represents segments of A- and B-block respectively.**

**Figure 6. (a) Change in fractional crystallinity ($X_c$) with $U_p$ at $\lambda$ = 1 for different block compositions ($x_B$) (b) Change in saturation crystallinity ($X$) at $U_p$ = 0.6 with block compositions ($x_B$). The lines joining the points are meant only as a guide to the eye.**

**Figure 7. (a) Change in fractional crystallinity ($X_c$) with $U_p$ at $\lambda$ = 4 for different block compositions ($x_B$); (b) Change in saturation crystallinity ($X$) at $U_p$ = 0.6 with block compositions ($x_B$). The lines joining the points are meant only as a guide to the eye.**

**Figure 8. Snapshots of semi crystalline structures at $\lambda$ = 1 for (a) $x_B$ = 0.25 and (b) $x_B$ = 0.75 during non-isothermal crystallization. Blue and orange lines represent crystalline bonds of A- and B-block respectively; yellow lines represent non-crystalline bonds of both the blocks.**

**Figure 9. Snapshots of semi crystalline structures at $\lambda$ = 4 for $x_B$) of (a) $x_B$ = 0.25 and (b) $x_B$ = 0.75 during non-isothermal crystallization. Blue and orange lines represent crystalline bonds of A- and B-block respectively; yellow lines represent non-crystalline bonds of both the blocks.**

**Figure 10. Change in (a) average crystallites size $\langle S \rangle$ and (b) average lamellar thickness $\langle l \rangle$ of A- and B-blocks with block compositions ($x_B$) at $\lambda = 1$. The lines joining the points are meant only as a guide to the eye.**

**Figure 11. Change in (a) average crystallites size $\langle S \rangle$ and (b) average lamellar thickness $\langle l \rangle$ with block compositions ($x_B$) at $\lambda = 4$. The lines joining the points are meant only as a guide to the eye.**

**Figure 12. Change in Overall Crystallinity ($X_c$) with Monte Carlo steps ($MCS$) on isothermal one step cooling at (a) $\lambda = 1$ and (b) $\lambda = 4$. The lines joining the points are meant only as a guide to the eye.**

**Figure 13. Change in Avrami index $(n)$ with block composition ($x_B$) for A- and B-block at (a) $\lambda = 1$ and (b) $\lambda = 4$. The lines joining the points are meant only as a guide to the eye.**

**Figure 14. Snapshots of semi-crystalline structures for $x_B = 0.25$ at $\lambda = 1$ during (a) isothermal two-step cooling at $U_p = 0.3$, (b) isothermal two-step cooling at $U_p = 0.6$, and (c) isothermal one-step cooling at $U_p = 0.6$. Blue and orange lines represent crystalline bonds of A- and B-block respectively; yellow lines represent non-crystalline bonds of both the blocks.**

**Figure 15. Snapshots of semi-crystalline structures for $x_B = 0.75$ at $\lambda = 1$ during (a) isothermal two-step cooling at $U_p = 0.3$, (b) isothermal two-step cooling at $U_p = 0.6$, and (c) isothermal one-step cooling at $U_p = 0.6$. Blue and orange lines represent crystalline bonds of A- and B-block respectively; yellow lines represent non-crystalline bonds of both the blocks.**

**Table 1**

| Composition $(x_B)$ | Two-step cooling | | One-step cooling |
|---|---|---|---|
| | $U_p = 0.3$ | $U_p = 0.6$ | $U_p = 0.6$ |
| 0.125 | 0.69 | 0.72 | 0.64 |
| 0.25 | 0.68 | 0.71 | 0.60 |
| 0.375 | 0.68 | 0.71 | 0.58 |
| 0.5 | 0.66 | 0.71 | 0.57 |
| 0.625 | 0.67 | 0.71 | 0.57 |
| 0.75 | 0.69 | 0.73 | 0.58 |
| 0.875 | 0.72 | 0.78 | 0.58 |

**Table 2**

| Composition ($x_B$) | Two-step cooling | | One-step cooling |
|---|---|---|---|
| | $U_p = 0.3$ | $U_p = 0.6$ | $U_p = 0.6$ |
| 0.125 | 0.05 | 0.53 | 0.50 |
| 0.25 | 0.08 | 0.56 | 0.53 |
| 0.375 | 0.10 | 0.58 | 0.53 |
| 0.5 | 0.10 | 0.58 | 0.55 |
| 0.625 | 0.12 | 0.60 | 0.57 |
| 0.75 | 0.12 | 0.62 | 0.60 |
| 0.875 | 0.12 | 0.66 | 0.65 |

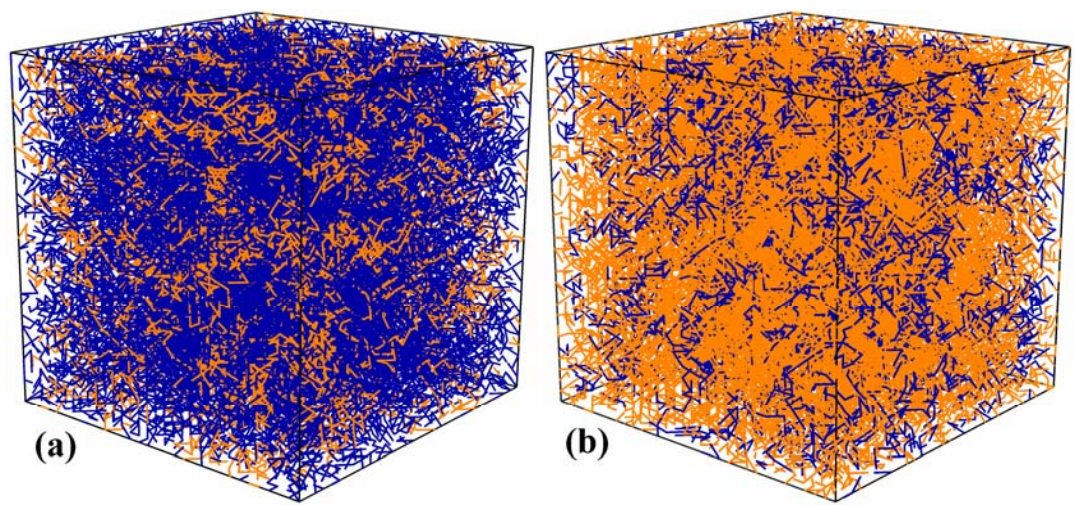

**Figure 1**

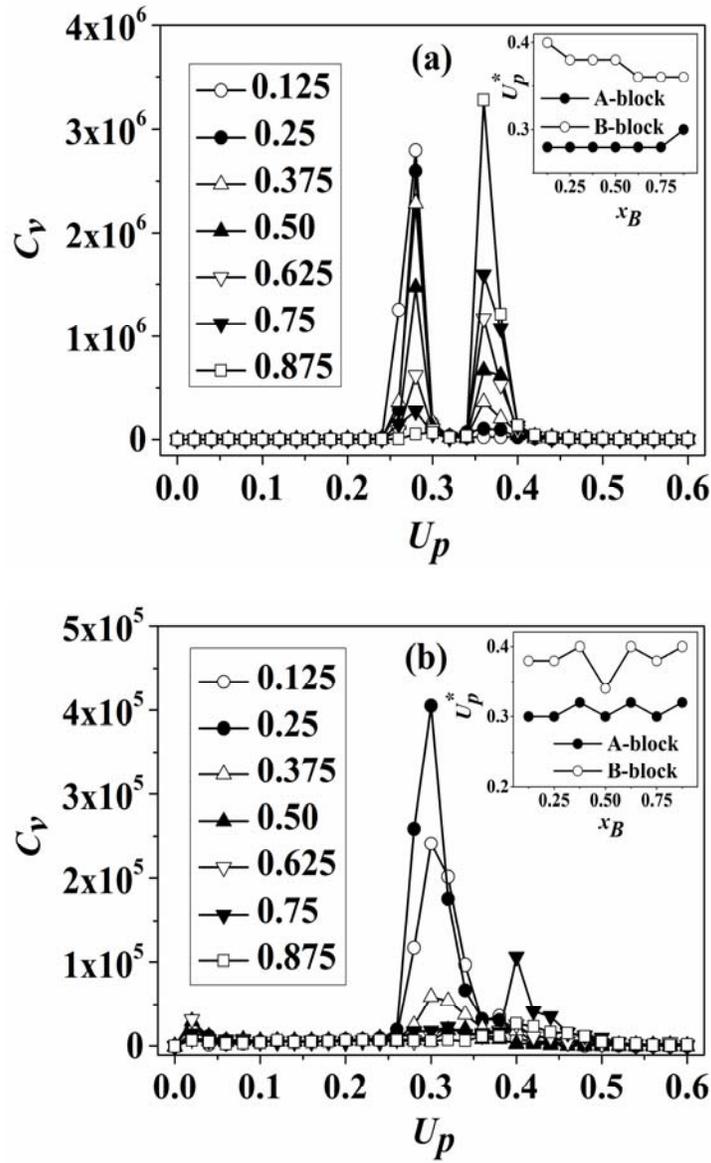

**Figure 2**

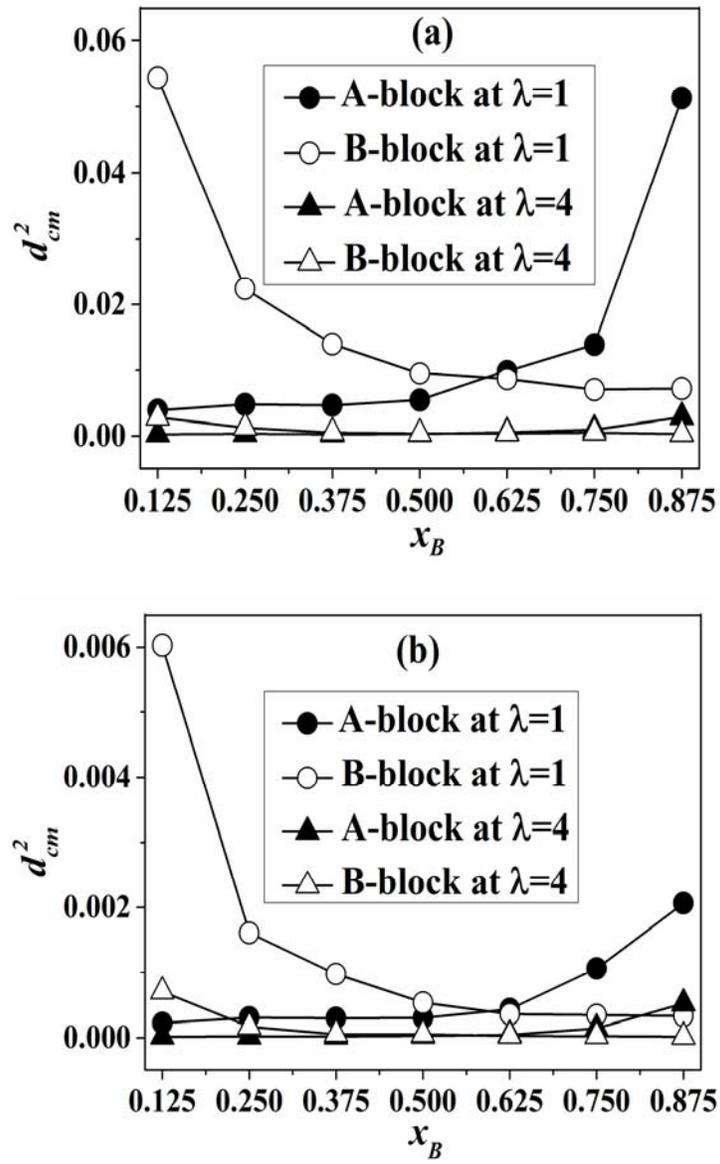

**Figure 3**

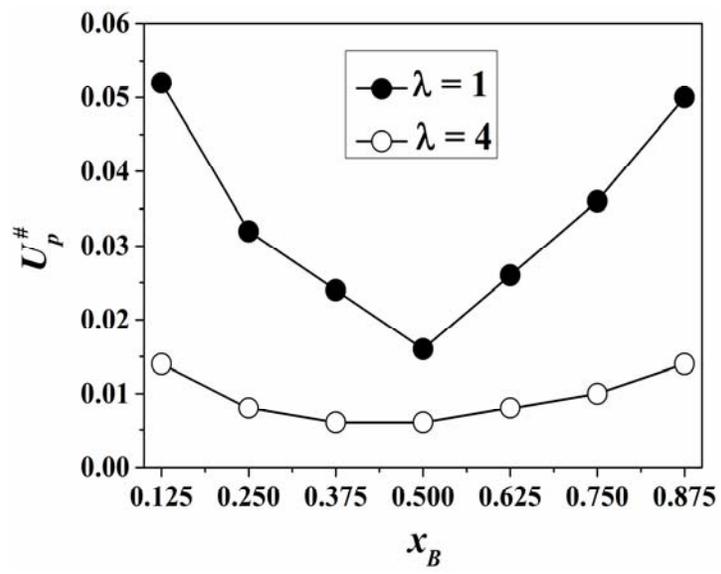

**Figure 4**

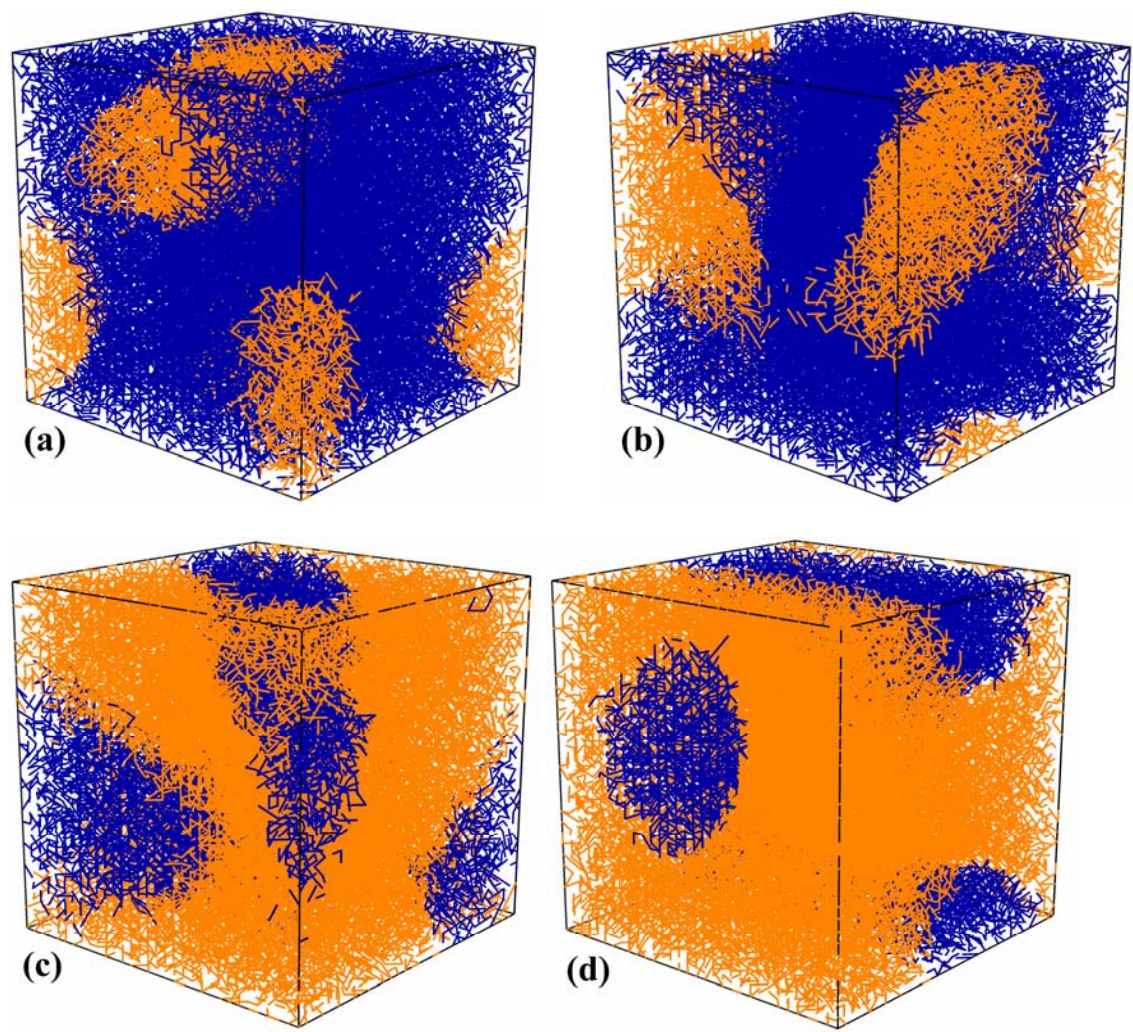

**Figure 5**

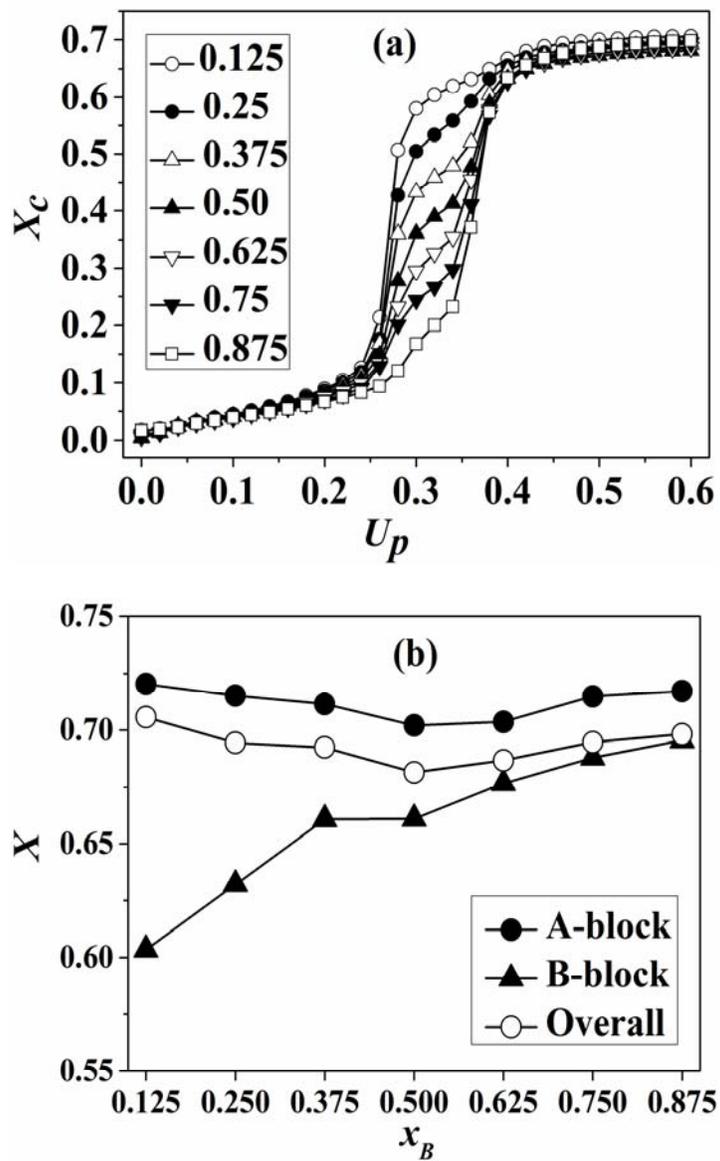

**Figure 6**

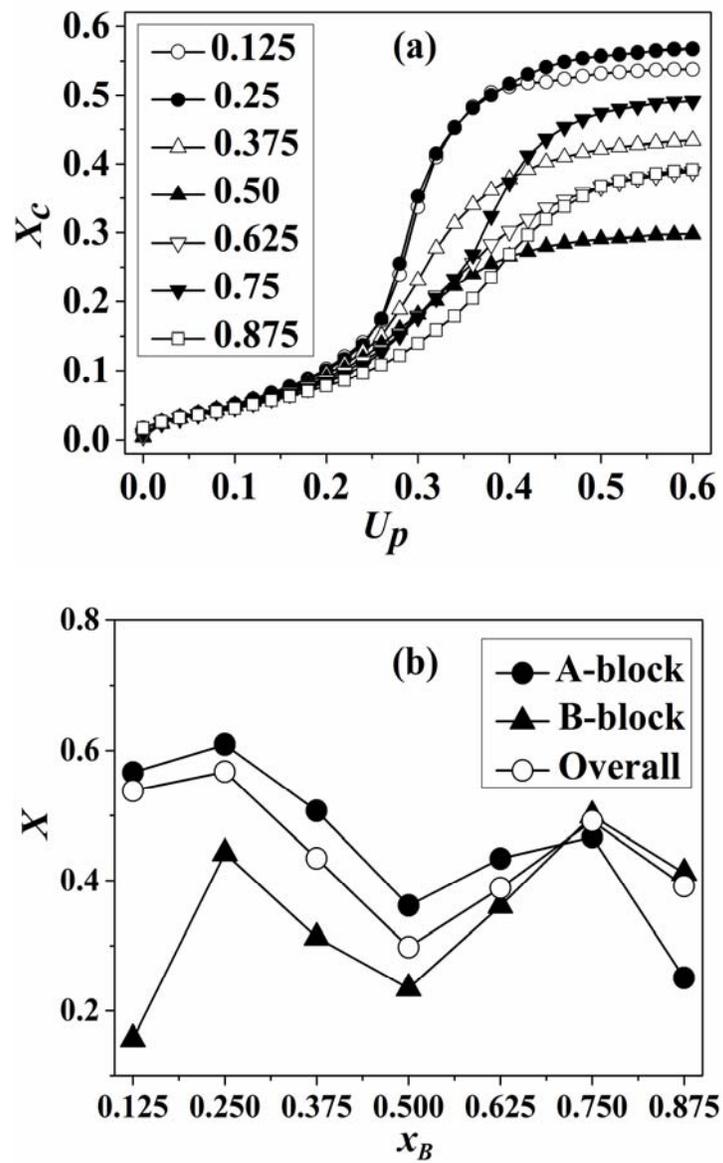

**Figure 7**

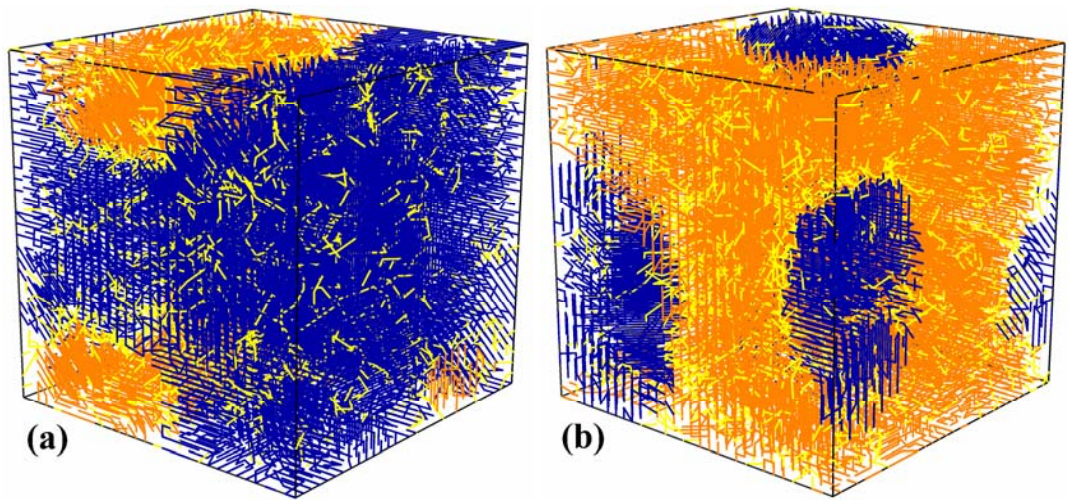

**(a)**

**(b)**

**Figure 8**

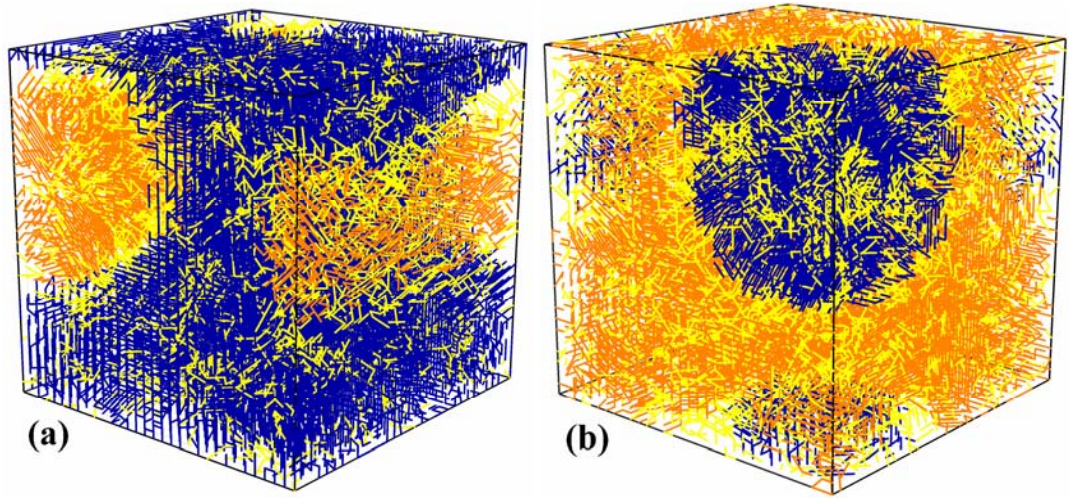

**Figure 9**

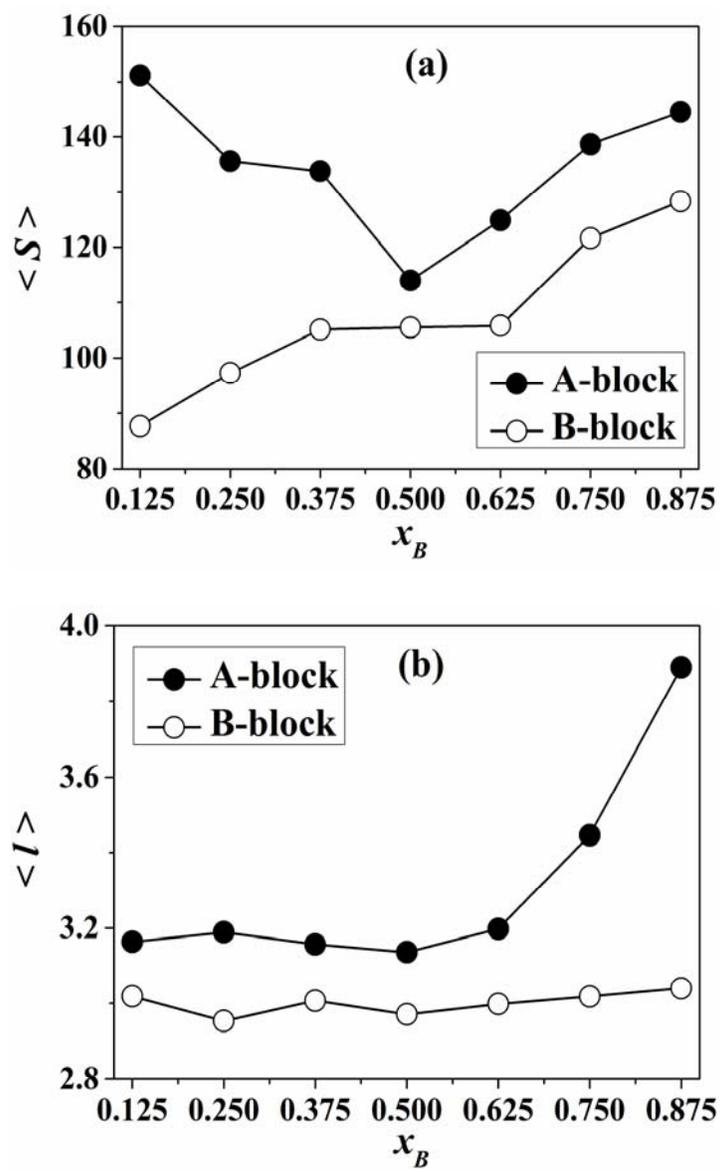

**Figure 10**

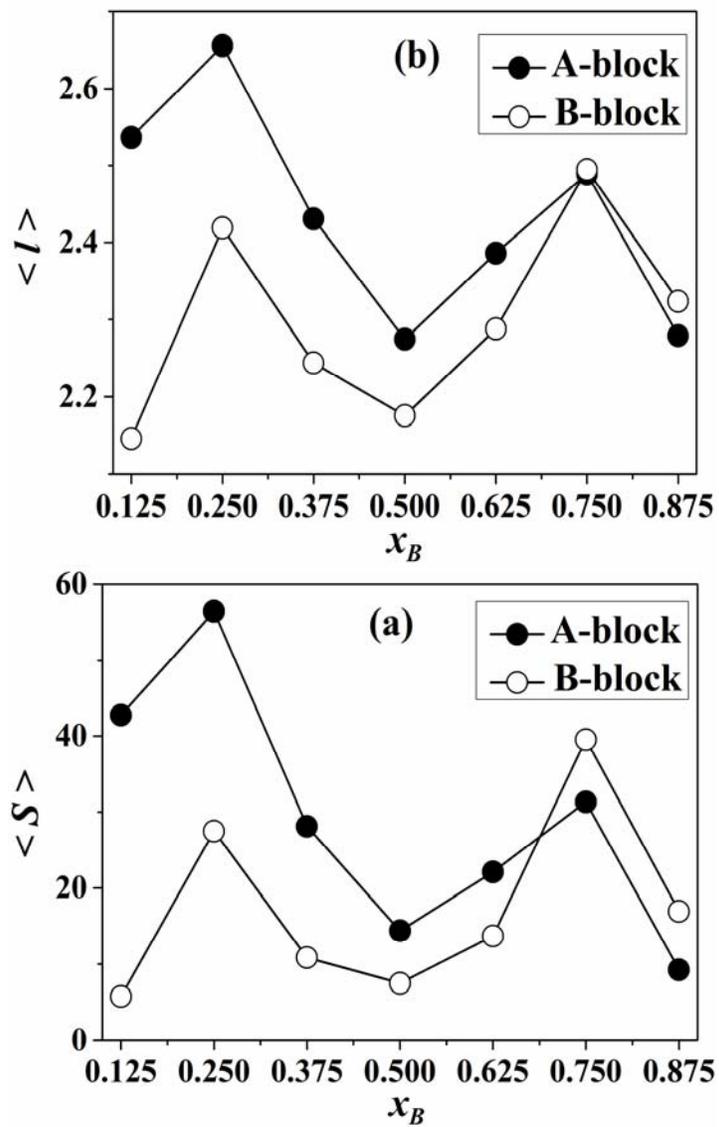

**Figure 11**

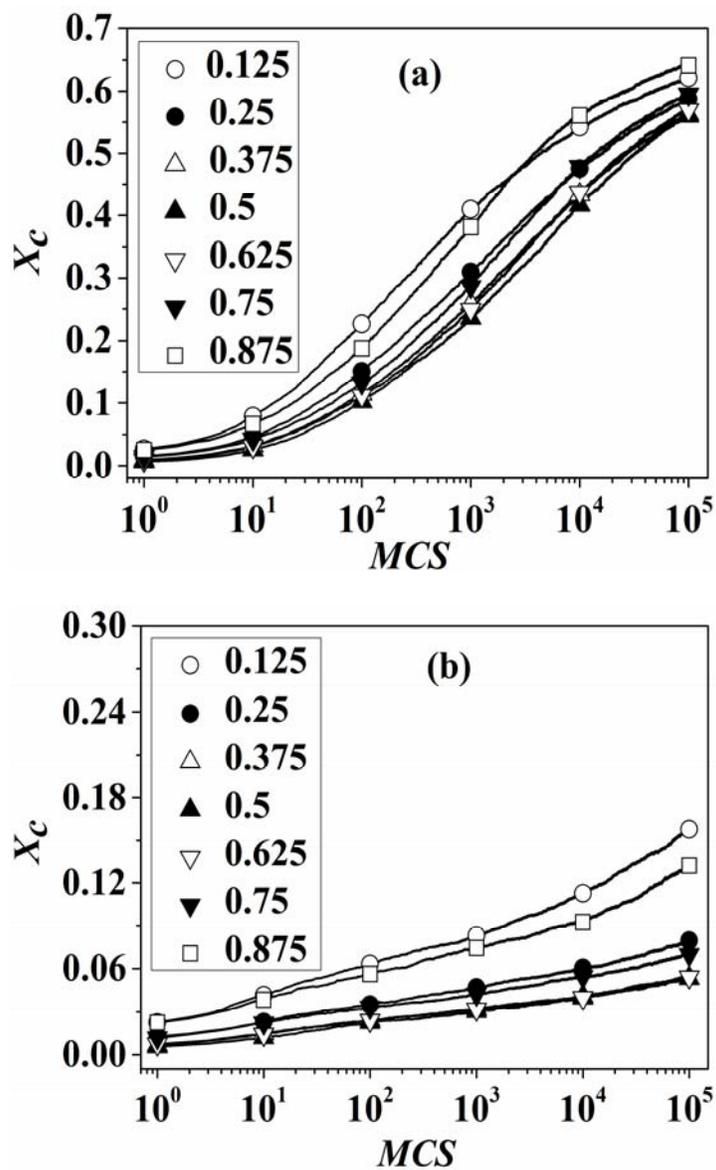

**Figure 12**

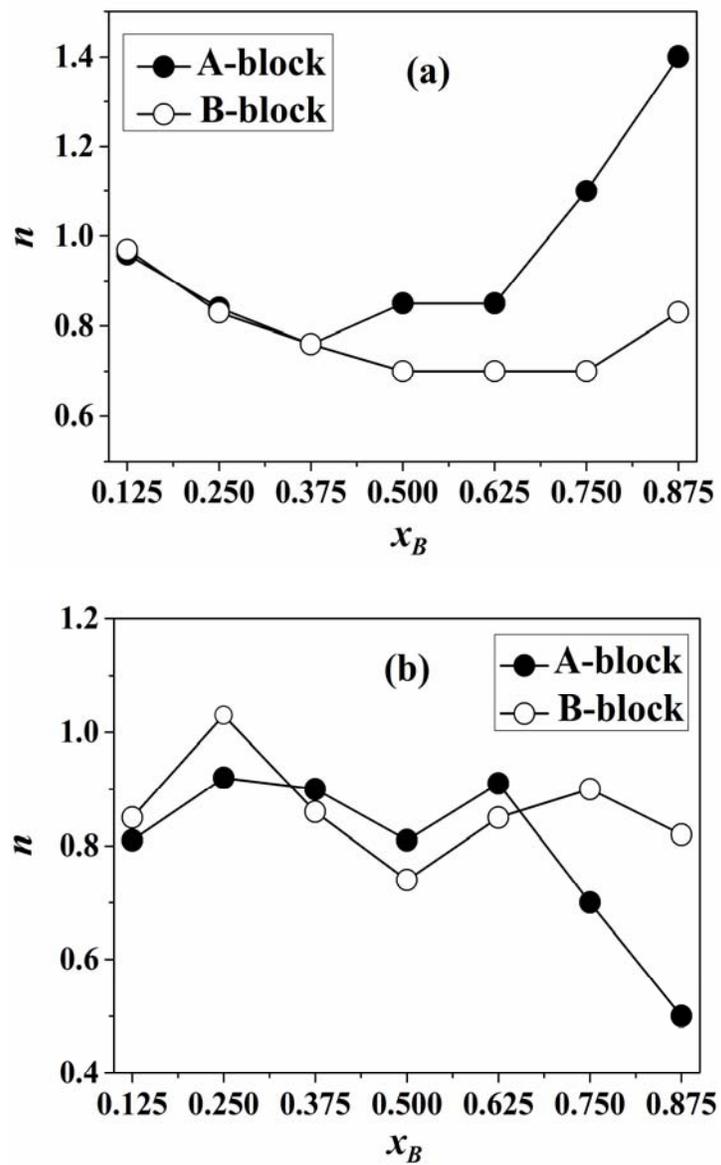

**Figure 13**

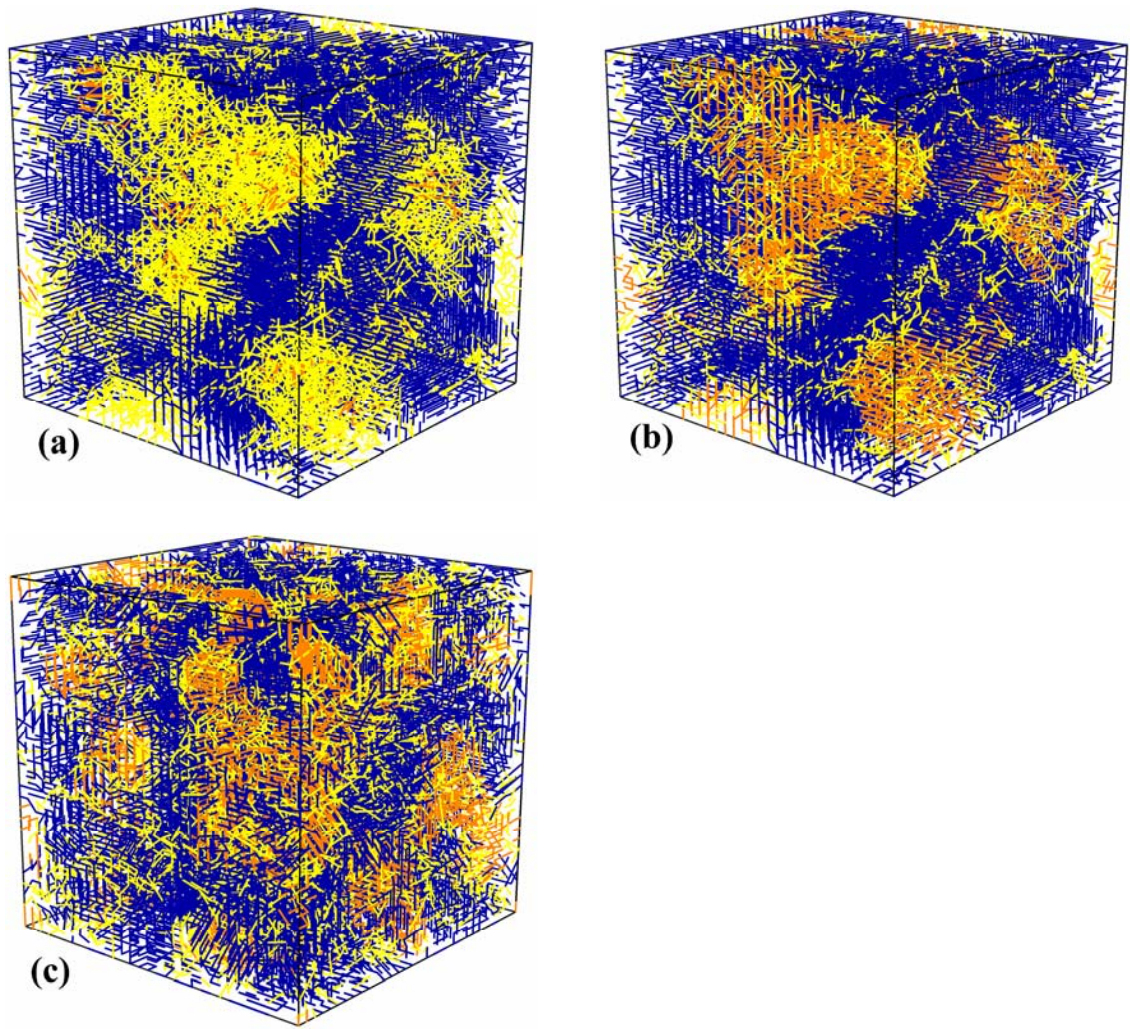

**Figure 14**

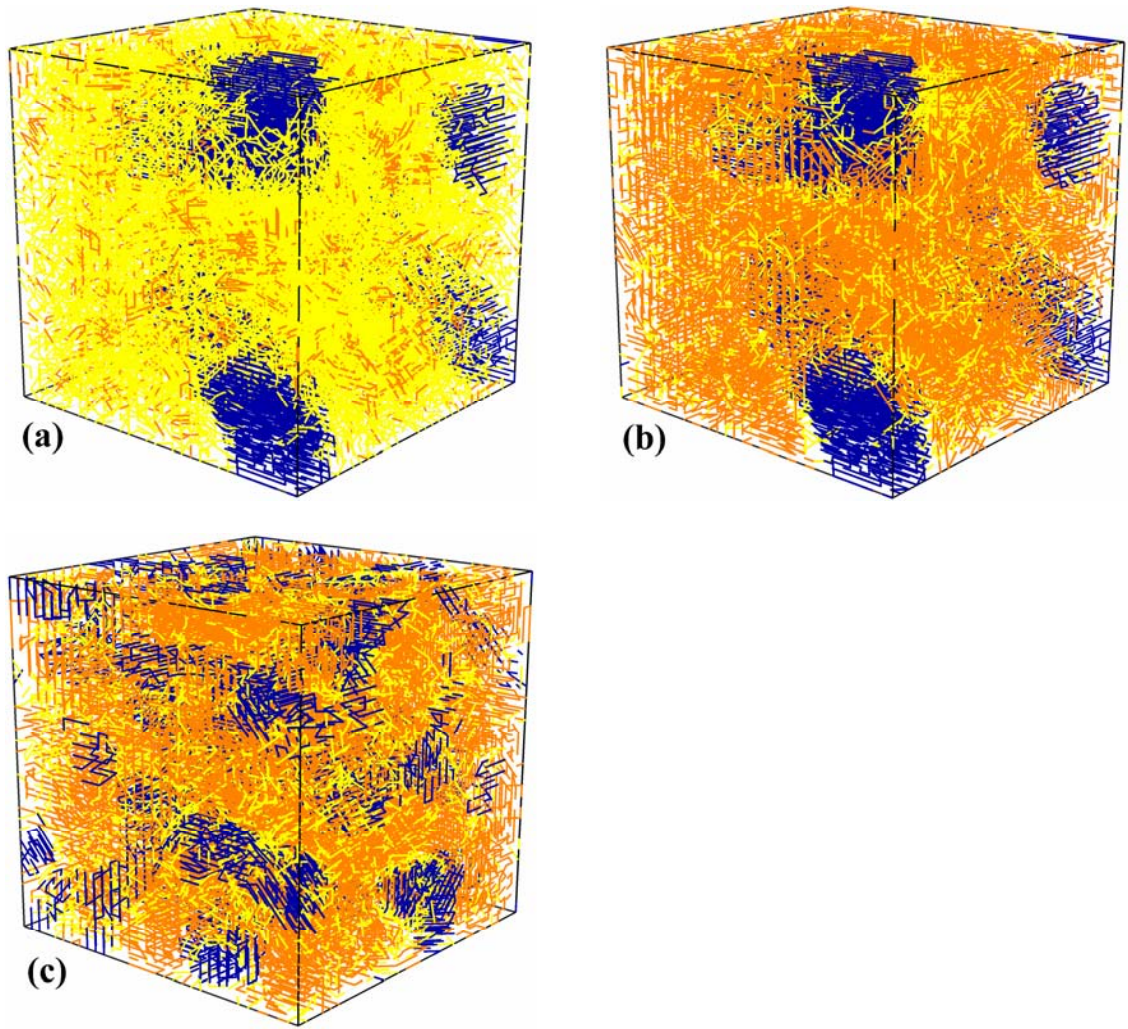

**(a)**

**(b)**

**(c)**

**Figure 15**

# Supporting Information for

## Effect of Block Asymmetry on the Crystallization of Double crystalline Diblock Copolymers

Chitrita Kundu and Ashok Kumar Dasmahapatra[*]

Department of Chemical Engineering, Indian Institute of Technology Guwahati, Guwahati – 781039, Assam, India

**Table S1. Comparison in Crystallinity of A-block ($X_A$) during two-step and one-step isothermal crystallization at $\lambda$ = 4. At strong segregation, the development of crystallinity in A-block is less ranges from 10 to 30% due to confinement created by large number of melt microdomains.**

| Composition ($x_B$) | Two-step cooling | | One-step cooling |
|---|---|---|---|
| | $U_p = 0.3$ | $U_p = 0.6$ | $U_p = 0.6$ |
| 0.125 | 0.33 | 0.35 | 0.175 |
| 0.25 | 0.190 | 0.20 | 0.09 |
| 0.375 | 0.150 | 0.155 | 0.06 |
| 0.50 | 0.130 | 0.135 | 0.06 |
| 0.625 | 0.116 | 0.120 | 0.05 |
| 0.75 | 0.109 | 0.114 | 0.04 |
| 0.875 | 0.106 | 0.112 | 0.04 |

[*] Corresponding author: Phone: +91-361-258-2273; Fax: +91-361-258-2291; Email address: akdm@iitg.ernet.in

**Table S2. Comparison in Crystallinity of B-block ($X_B$) during two-step and one-step isothermal crystallization at $\lambda = 4$. At strong segregation, the development of crystallinity in B-block is very less ranges up to 10 % as previously crystallize A-block creates confinement for the crystallization of B-block resulting slowing down of crystallinity of B-block.**

| Composition ($x_B$) | Two-step cooling | | One-step cooling |
|:---:|:---:|:---:|:---:|
| | $U_p = 0.3$ | $U_p = 0.6$ | $U_p = 0.6$ |
| 0.125 | 0.050 | 0.056 | 0.033 |
| 0.25 | 0.070 | 0.073 | 0.036 |
| 0.375 | 0.077 | 0.080 | 0.036 |
| 0.50 | 0.085 | 0.088 | 0.045 |
| 0.625 | 0.10 | 0.103 | 0.057 |
| 0.75 | 0.122 | 0.127 | 0.08 |
| 0.875 | 0.131 | 0.150 | 0.14 |

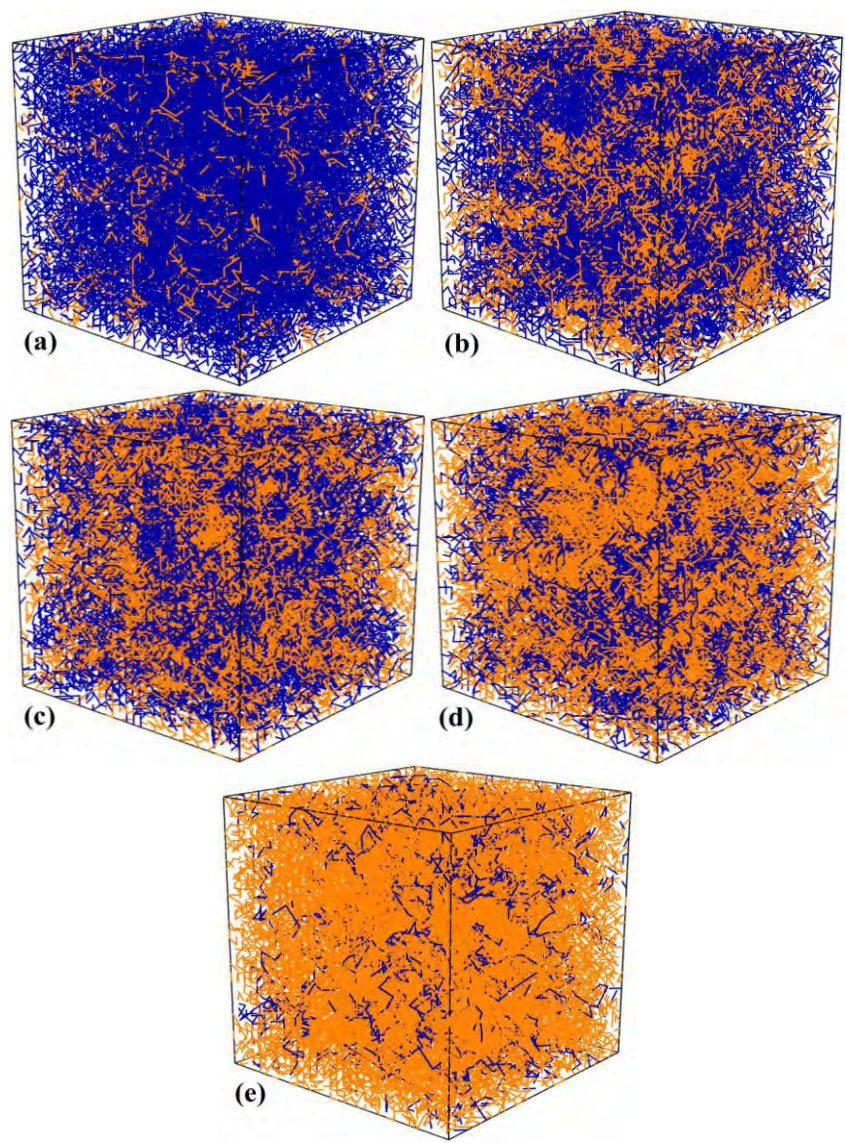

**Figure S1. Snapshots of homogeneous melt at** $U_p$ **= 0 for** $x_B$ **= (a) 0.125, (b) 0.375, (c) 0.50, (d) 0.625 and (e) 0.875. Blue and orange lines represent segments of A- and B-block respectively. The isotropic orientation of polymer chains between two blocks remains intact with the increment of B-units.**

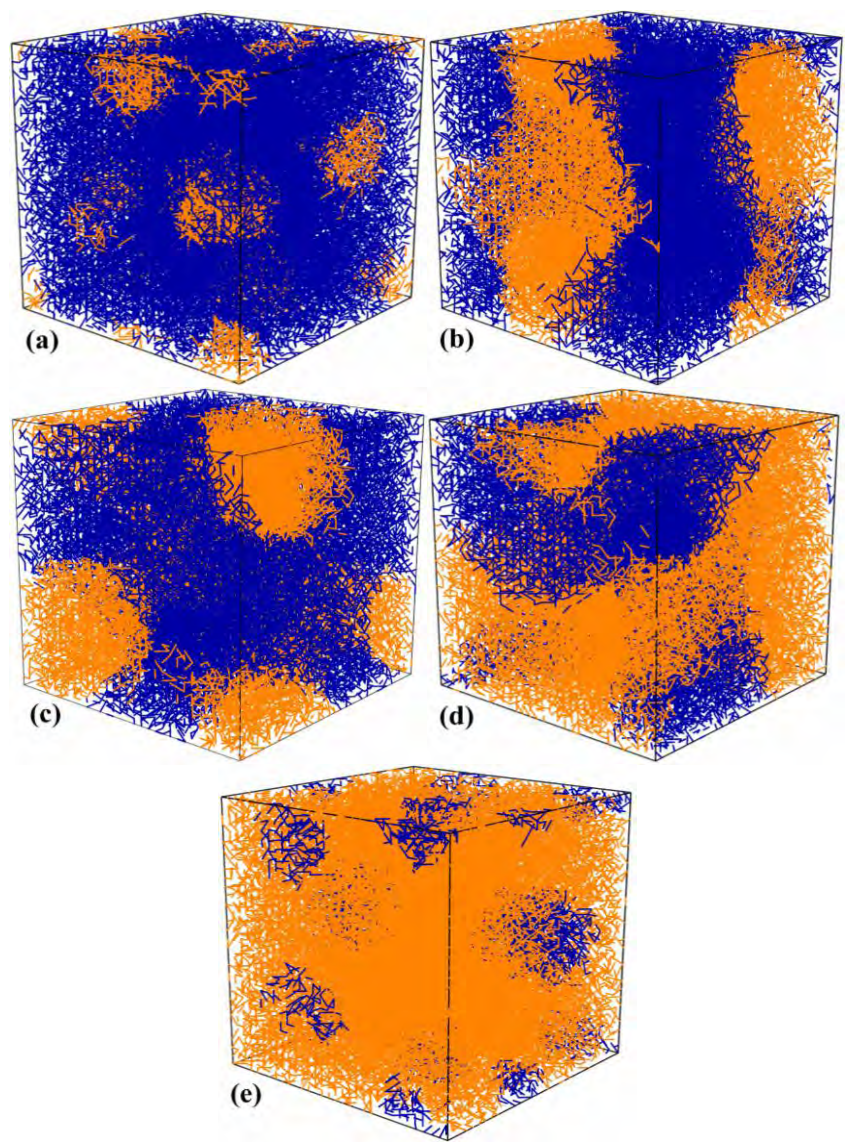

**Figure S2. Snapshots of microphase separated melt at** $U_p$ = **0.1 for** $x_B$ = **(a) 0.125, (b) 0.375, (c) 0.50, (d) 0.625 and (e) 0.875 at** $\lambda$ = **1. Blue and orange lines represent segments of A- and B-block respectively. The phase separated microdomains are clearly visible even at low segregation.**

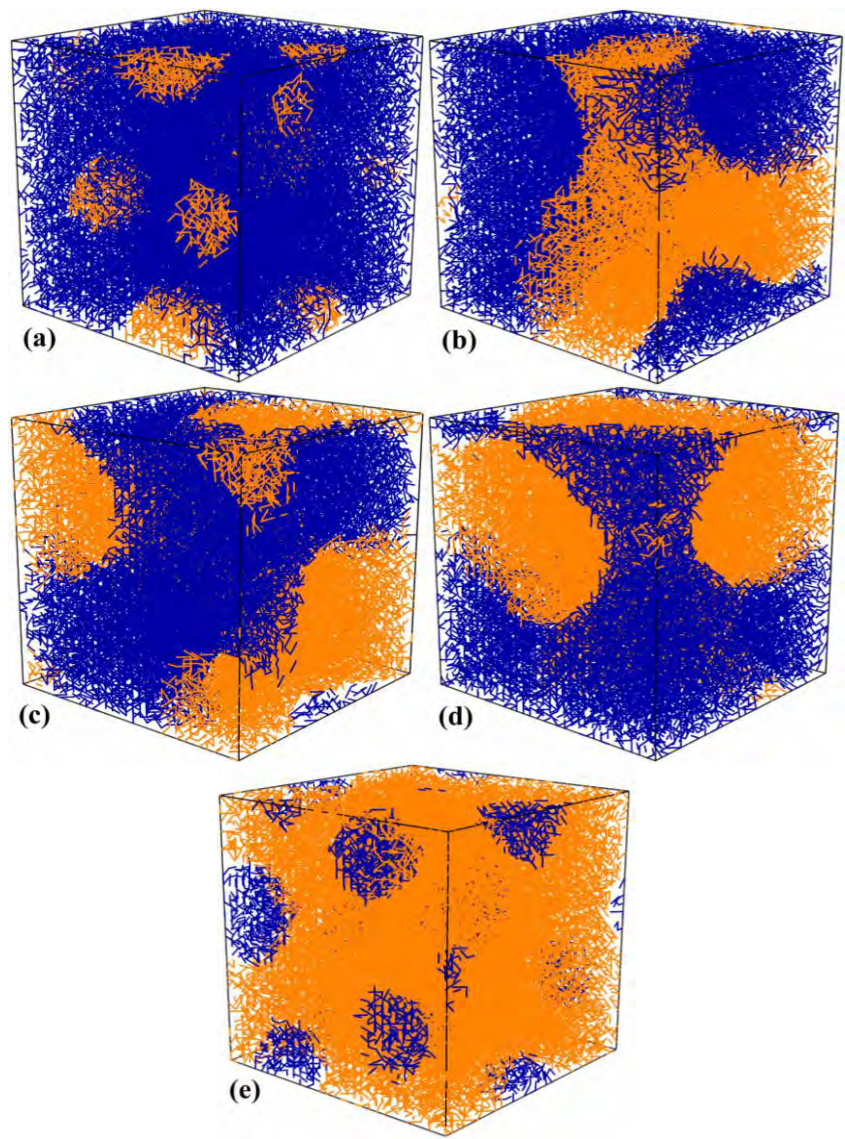

**Figure S3. Snapshots of microphase separated melt at** $U_p$ **= 0.1 for** $x_B$ **= (a) 0.125, (b) 0.375, (c) 0.50, (d) 0.625 and (e) 0.875 at** $\lambda$ **= 4. Blue and orange lines represent segments of A- and B-block respectively. Due to strong segregation, phase separated microdomins are more prominent than weak segregation.**

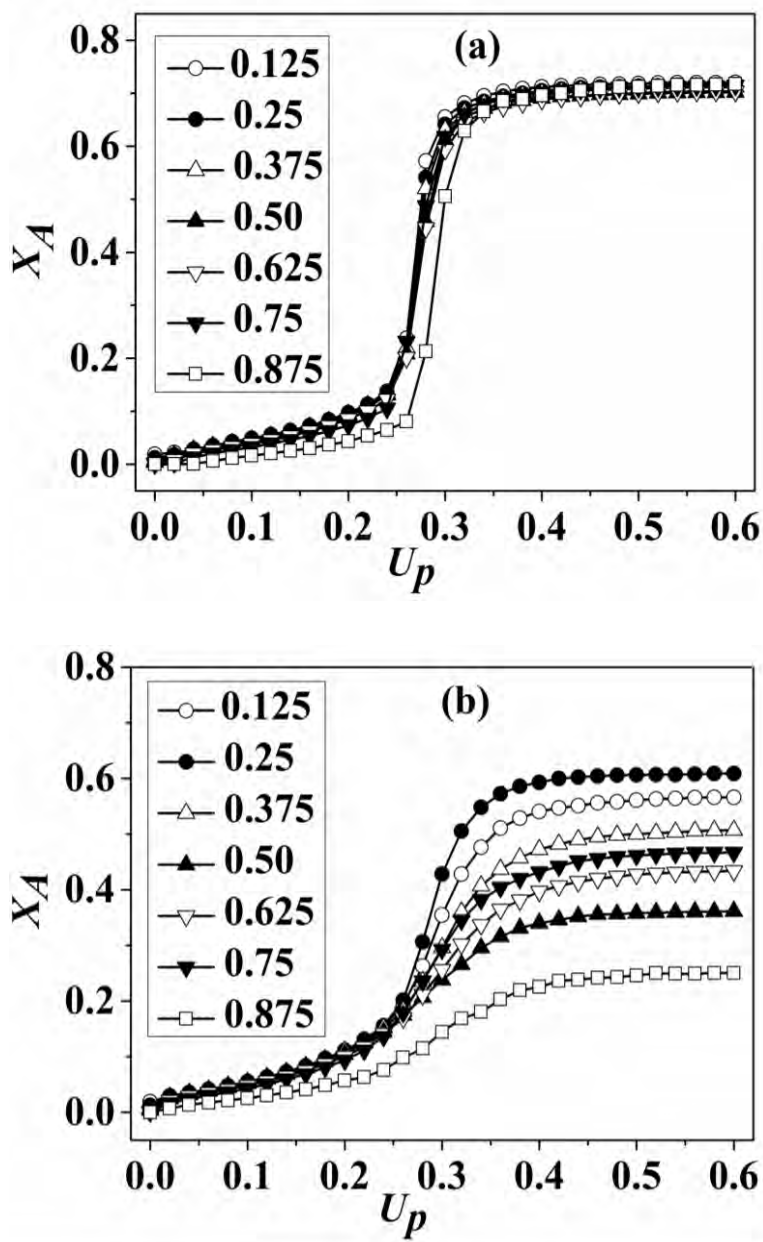

**Figure S4.** Change in crystallinity of A-block ( $X_A$ ) with $U_p$ of $x_B$ = 0.125, 0.25, 0.375, 0.5, 0.625, 0.75 and 0.875 at (a) $\lambda$ = 1 and (b) $\lambda$ = 4. The crystallinity ( $X_A$ ) increases with $U_p$, and at $U_p \sim$ 0.4, it reaches to a saturation crystallinity. The saturation crystallinity of A-block ( $X_A^{sat}$ ) remains same for all the block compositions ( $x_B$ ) at weak segregation but it gives non-monotonic trend with $x_B$ at strong segregation.

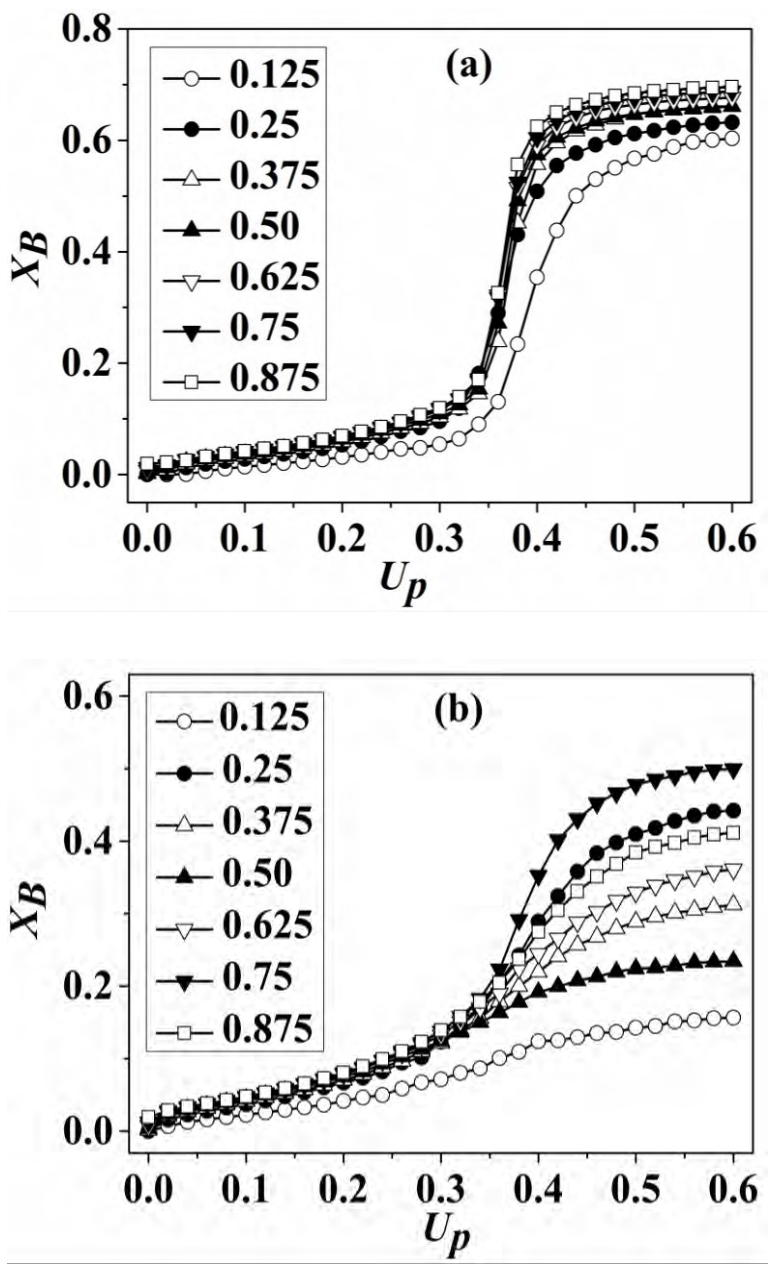

**Figure S5. Change in crystallinity of B-block ( $X_B$ ) with $U_p$ for $x_B$ = 0.125, 0.25, 0.375, 0.5, 0.625, 0.75 and 0.875 at (a) $\lambda$ = 1 and (b) $\lambda$ = 4. The crystallinity ( $X_B$ ) increases with $U_p$ and at $U_p \sim 0.5$, it reaches to saturation crystallinity ( $X_B^{sat}$ ). The saturation crystallinity of B-block ( $X_B^{sat}$ ) monotonically increases with $x_B$ at weak segregation whereas it gives non-monotonic trend with $x_B$ at strong segregation.**

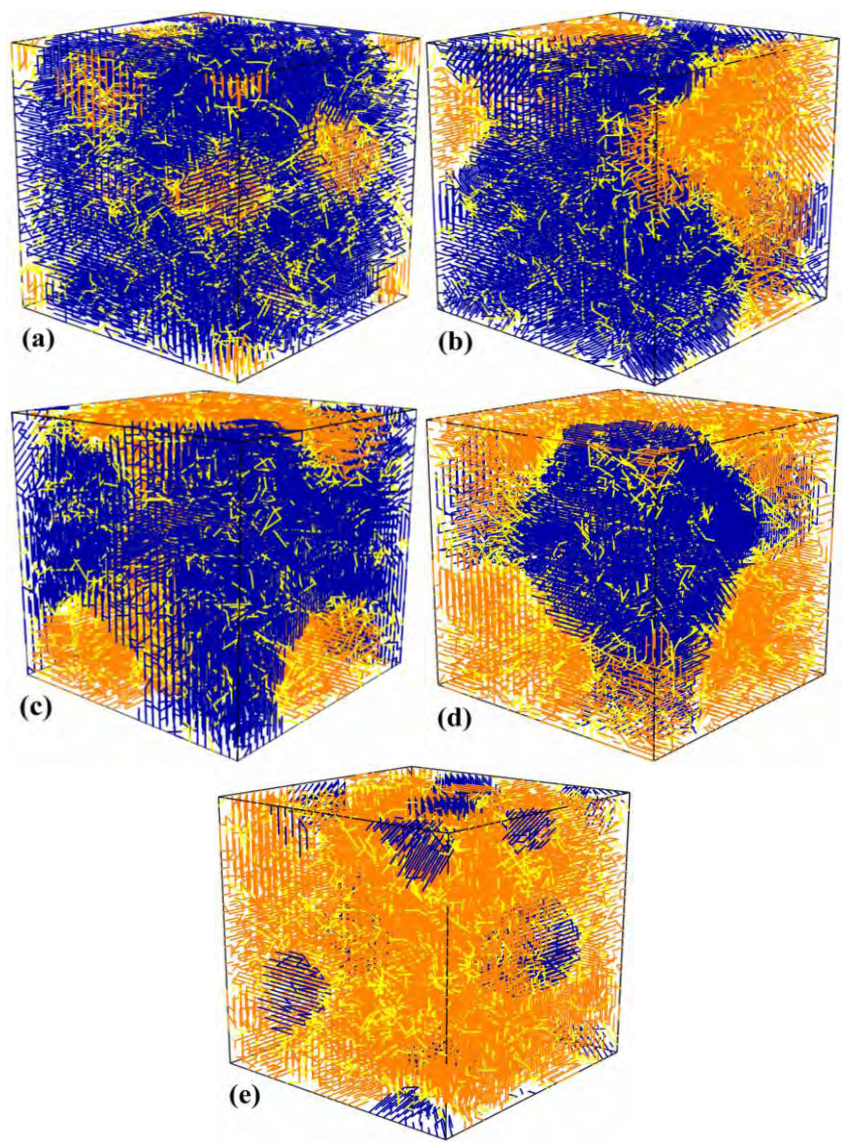

**Figure S6. Snapshots of semi crystalline structure of diblock copolymer at** $U_p$ **= 0.6 for** $x_B$ **= (a) 0.125, (b) 0.375, (c) 0.50, (d) 0.625 and (e) 0.875 at** $\lambda$ **= 1. Blue and orange lines represent crystalline bonds of A-block and B-block respectively, and yellow lines represent non-crystalline bonds of both the blocks.**

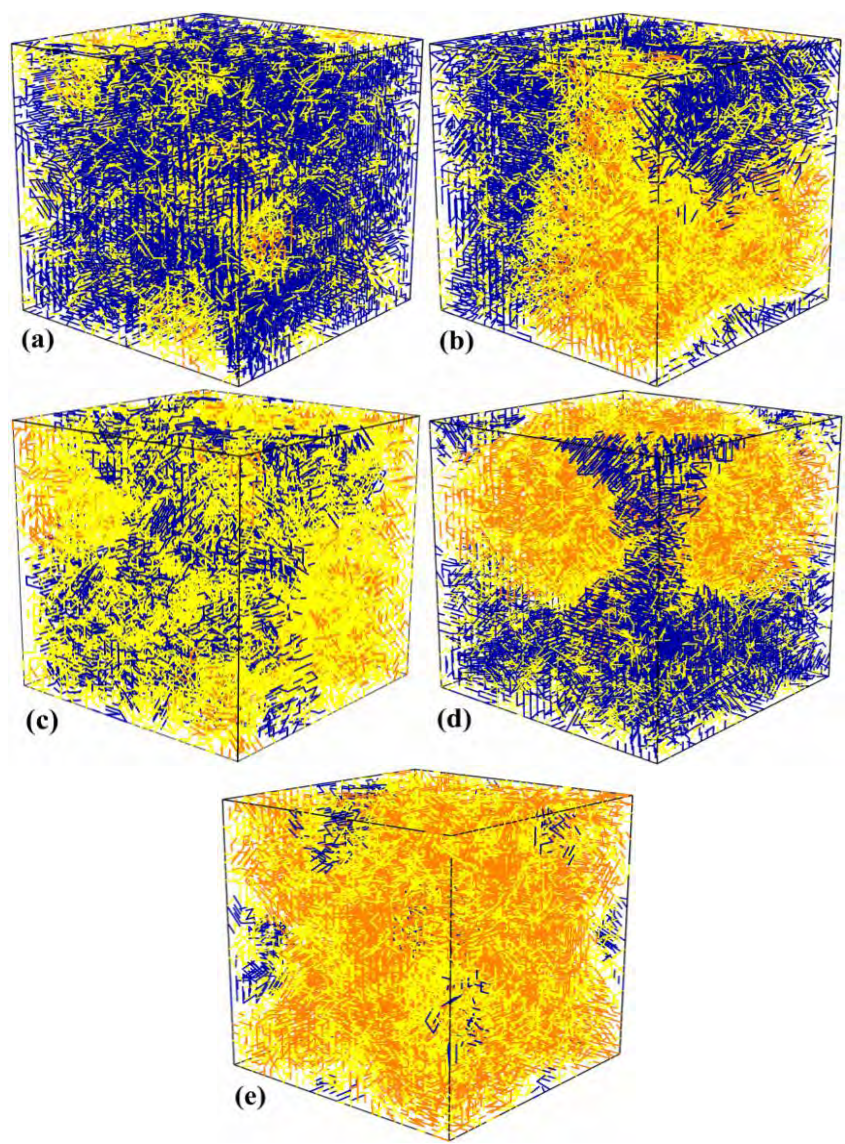

**Figure S7.** Snapshots of semi-crystalline structure of diblock copolymer at $U_p$ = 0.6 for $x_B$ = (a) 0.125, (b) 0.375, (c) 0.50, (d) 0.625 and (e) 0.875 at $\lambda$ =4. Blue and orange lines represent crystalline bonds of A-block and B-block respectively, and yellow lines represent non-crystalline bonds of both the blocks. Due to strong segregation, the amorphous layers of both the blocks are increased as the crystallinity of both the blocks is supressed by large number of microdomains in microphase separated melt.

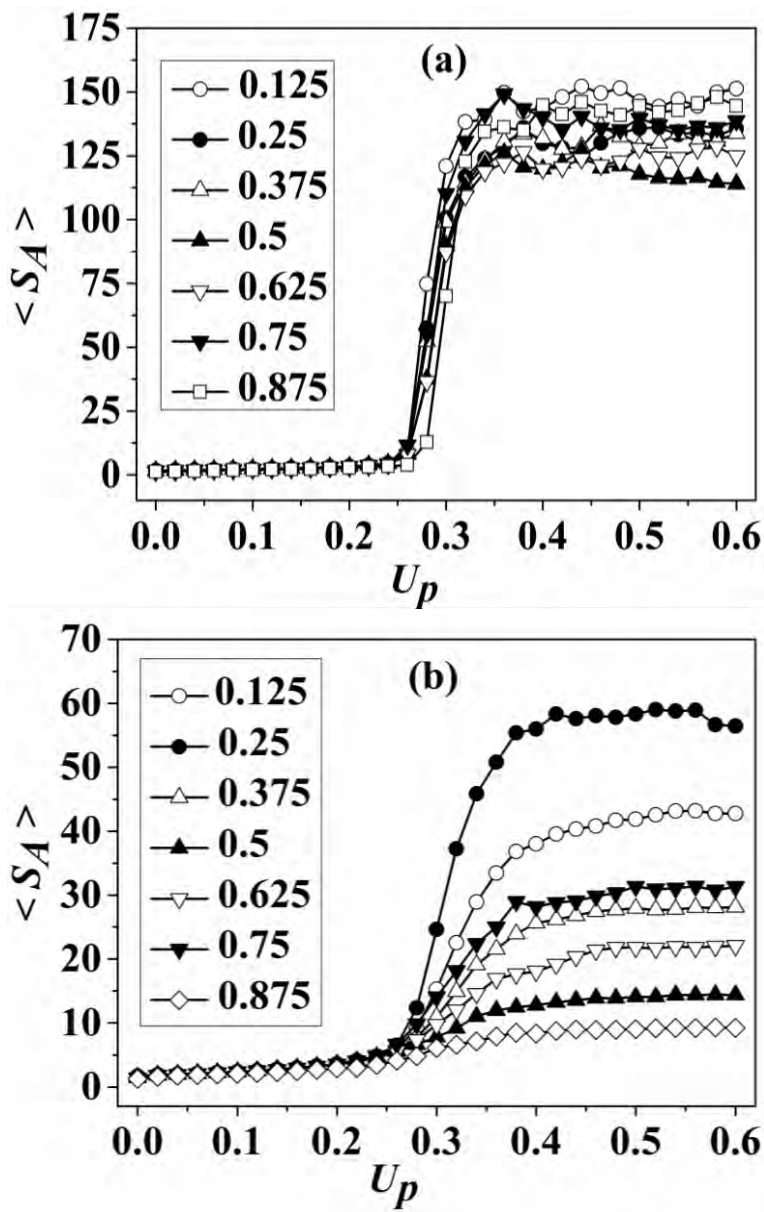

**Figure S8.** Change in average crystallites size of A-block $\langle S_A \rangle$ with $U_p$ for $x_B$ = **0.125,**

**0.25, 0.375, 0.5, 0.625, 0.75 and 0.875 at (a)** $\lambda$ **= 1 and (b)** $\lambda$ **= 4. A crystallite is a small**

**microscopic aggregate having crystalline bonds in the same orientation. The crystallite**

**size is measured as the total number of crystalline bonds present in it. With increased**

**value of** $U_p$, **average crystallites size of A-block is increased. There is a non-monotonic**

**trend of** $\langle S_A \rangle$ **with block composition (** $x_B$ **) in both the segregations.**

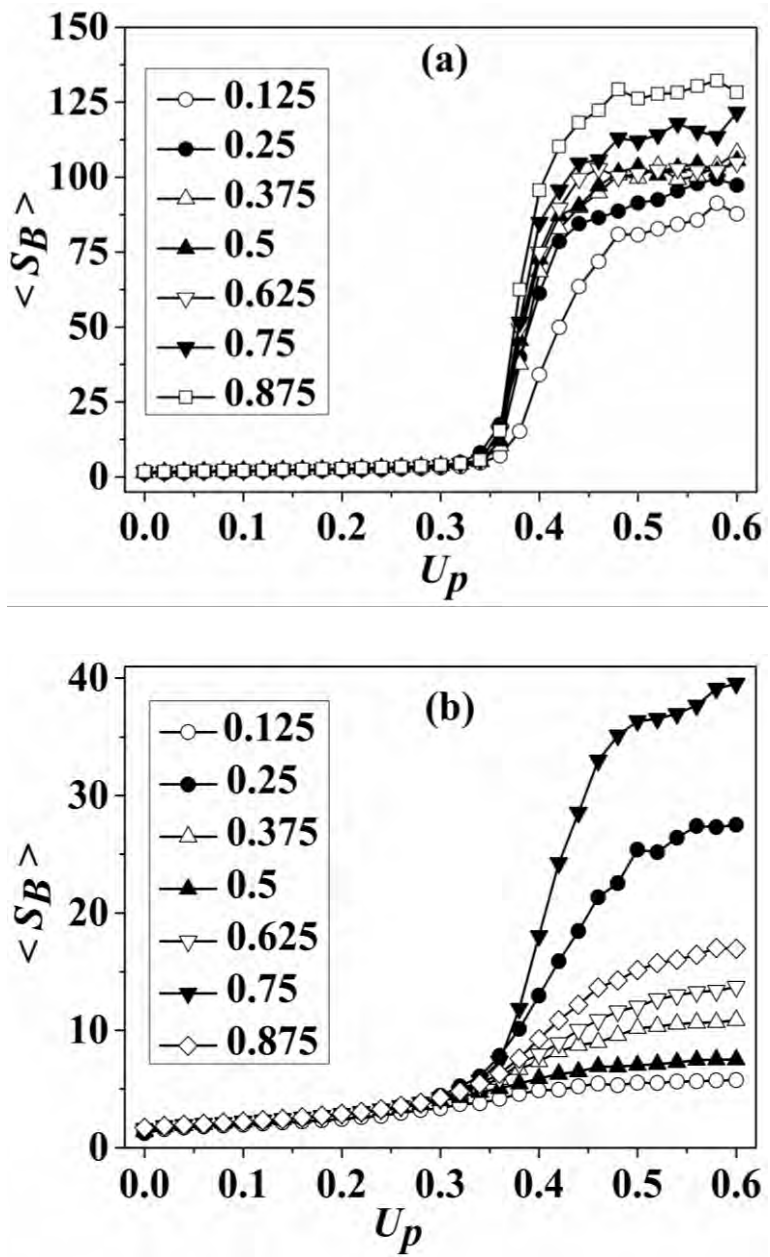

**Figure S9. Change in average crystallites size of B-block** $\langle S_B \rangle$ **with** $U_p$ **for** $x_B$ **= 0.125, 0.25, 0.375, 0.5, 0.625, 0.75 and 0.875 at (a)** $\lambda$ **= 1 and (b)** $\lambda$ **= 4. With increased value of** $U_p$ **, average crystallites size of B-block is also increased. At weak segregation, there is a monotonic increase in** $\langle S_B \rangle$ **with block composition (** $x_B$ **) whereas at strong segregation, it follows a non-monotonic trend.**

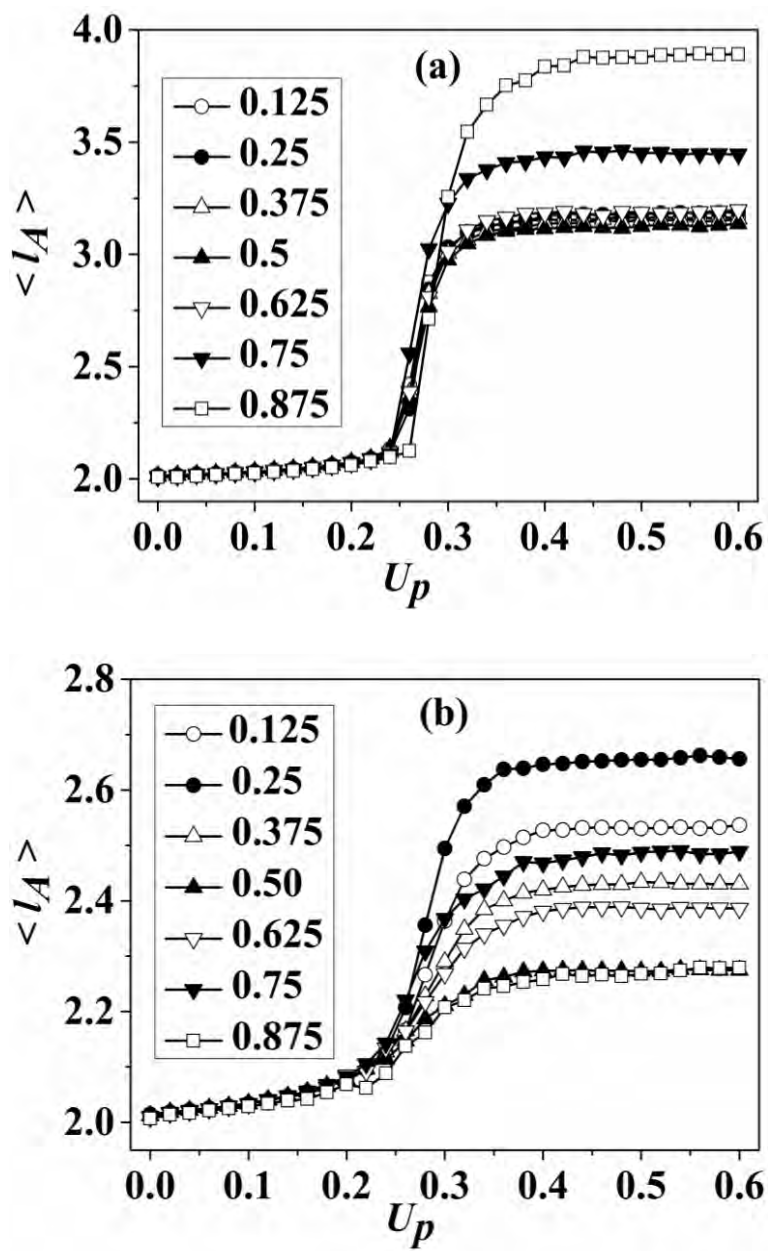

**Figure S10. Change in average lamellar thickness of A-block $\langle l_A \rangle$ with $U_p$ for $x_B =$ 0.125, 0.25, 0.375, 0.5, 0.625, 0.75 and 0.875 at (a) $\lambda = 1$ and (b) $\lambda = 4$. The lamellar thickness is the average number of block units towards the direction of crystal thickness in a given crystallite, and their average is calculated over all the crystallites present in the system. With increased value of $U_p$, $\langle l_A \rangle$ increases. At weak segregation, $\langle l_A \rangle$ remains same for most of the block composition ($x_B$) whereas for $x_B = 0.75$ and 0.875, it suddenly increases, which is attributed to the dilution effect imposed by B-block during crystallization. At strong segregation, it again follows a non-monotonic trend.**

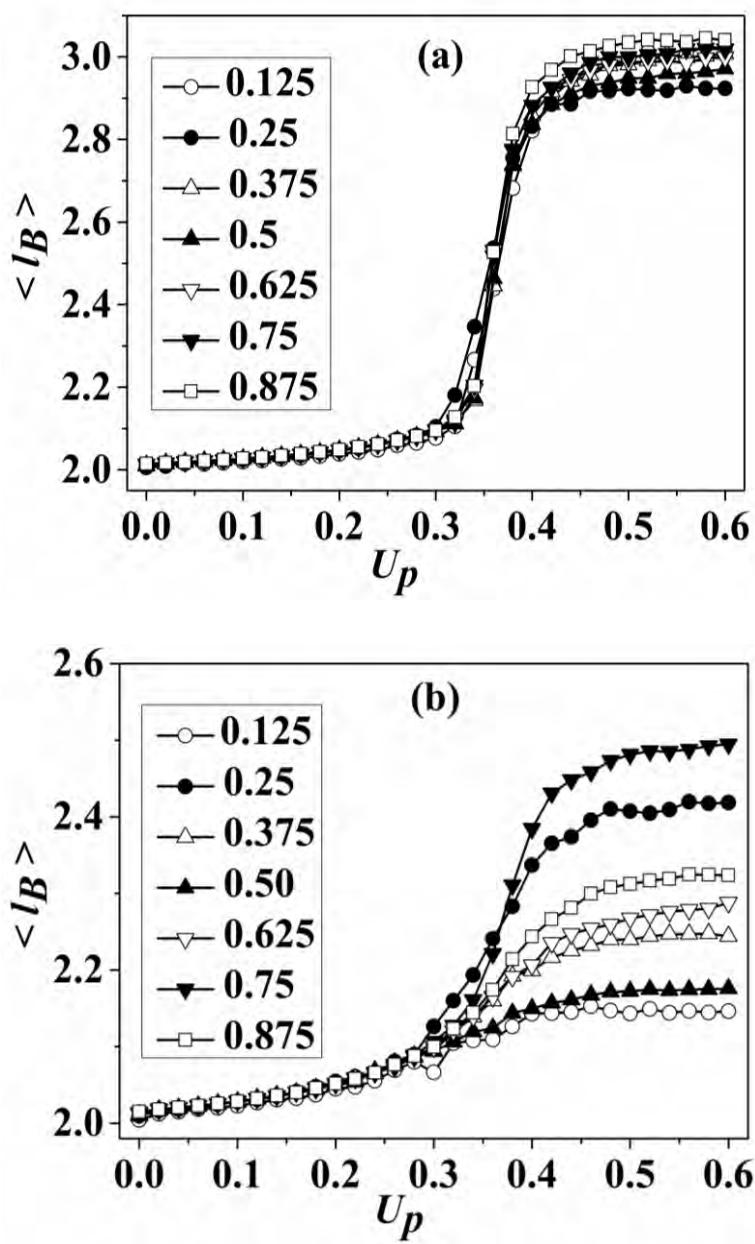

**Figure S11. Change in average lamellar thickness of B-block** $\langle l_B \rangle$ **with** $U_p$ **for** $x_B =$ **0.125, 0.25, 0.375, 0.5, 0.625, 0.75 and 0.875 at (a)** $\lambda = 1$ **(b)** $\lambda = 4$**. With increased value of** $U_p$**,** $\langle l_B \rangle$ **increases. There is a monotonic increase in** $\langle l_B \rangle$ **with block composition of B (** $x_B$ **) at weak segregation, whereas it follows a non-monotonic trend at strong segregation.**

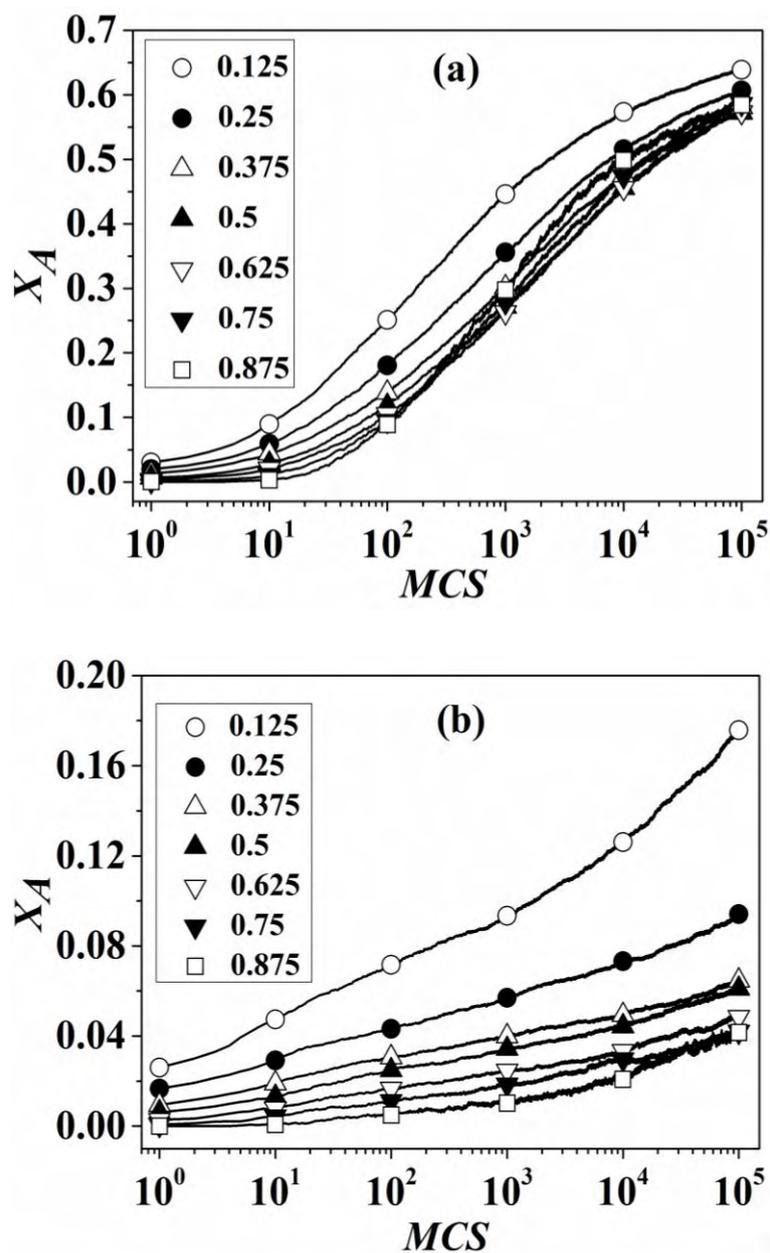

**Figure S12.** **Development in crystallinity of A-block (** $X_A$ **) during one-step isothermal crystallization at (a)** $\lambda$ **= 1 and (b)** $\lambda$ **= 4. At weak segregation, there is a non-monotonic trend of crystallinity of A-block (** $X_A$ **) with block composition (** $x_B$ **) but at strong segregation it follows a monotonic trend with block composition (** $x_B$ **).**

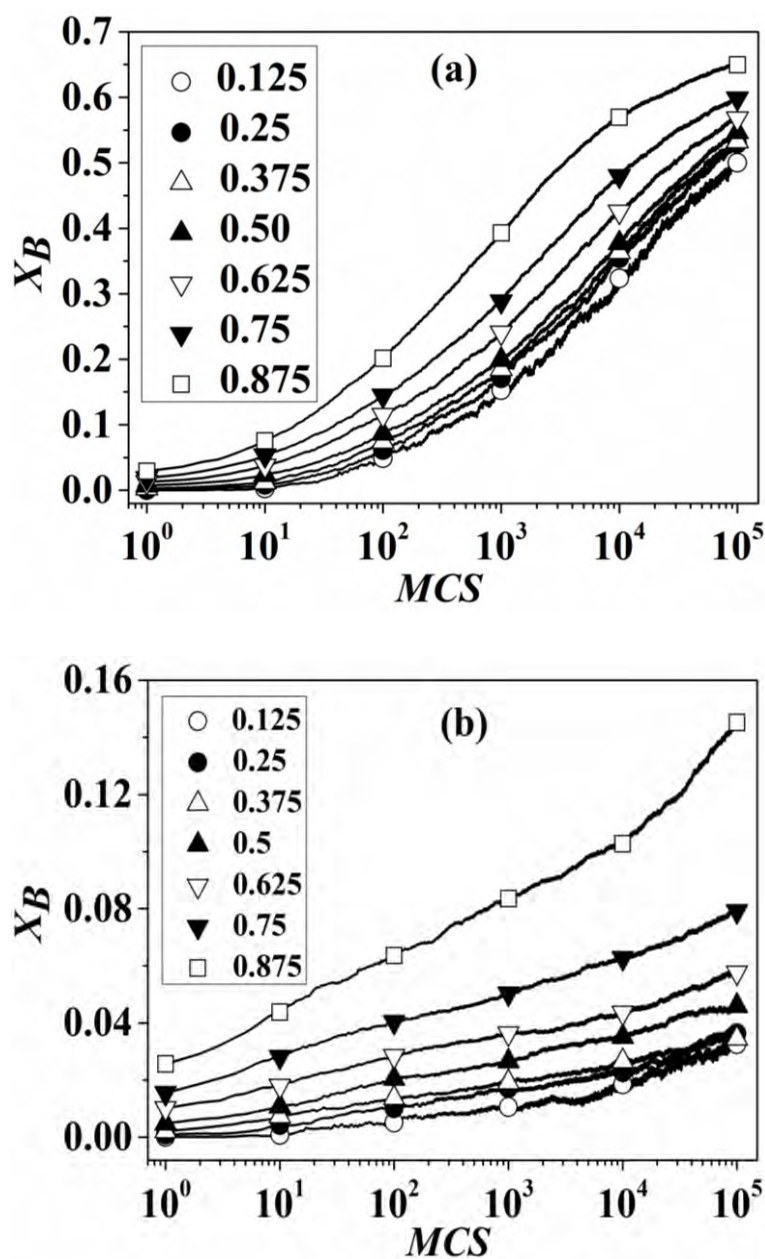

**Figure S13.** Development in crystallinity of B-block ( $X_B$ ) during one-step isothermal crystallization at (a) $\lambda = 1$ and (b) $\lambda = 4$. There is a monotonic increase in crystallinity of B-block ( $X_B$ ) with block composition ( $x_B$ ) at both segregation.

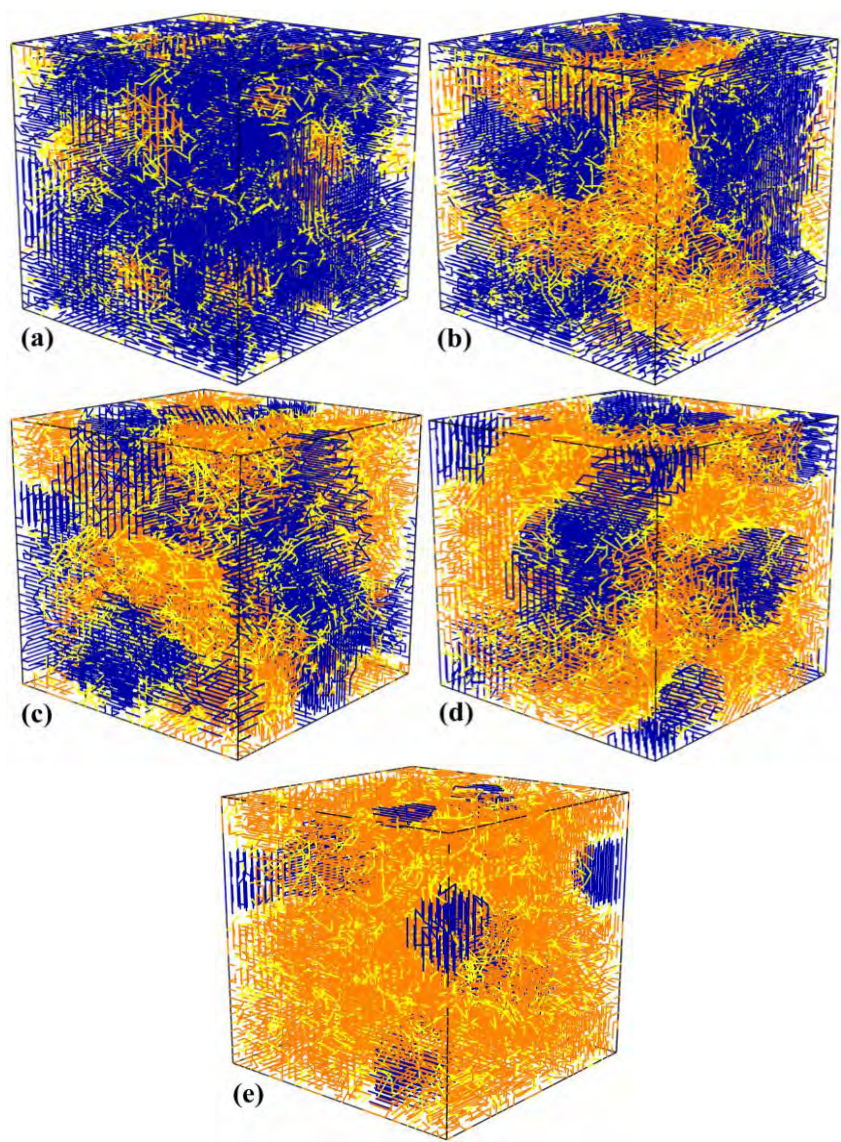

**Figure S14.** Snapshots of semi crystalline structure of diblock copolymer at $U_p$ = 0.6 for $x_B$ = **(a) 0.125, (b) 0.375, (c) 0.50, (d) 0.625 and (e) 0.875 during two-step isothermal crystallization at** $\lambda$ **= 1. Blue and orange lines represent crystalline bonds of A-block and B-block respectively, and yellow lines represent non-crystalline bonds of both the blocks.**

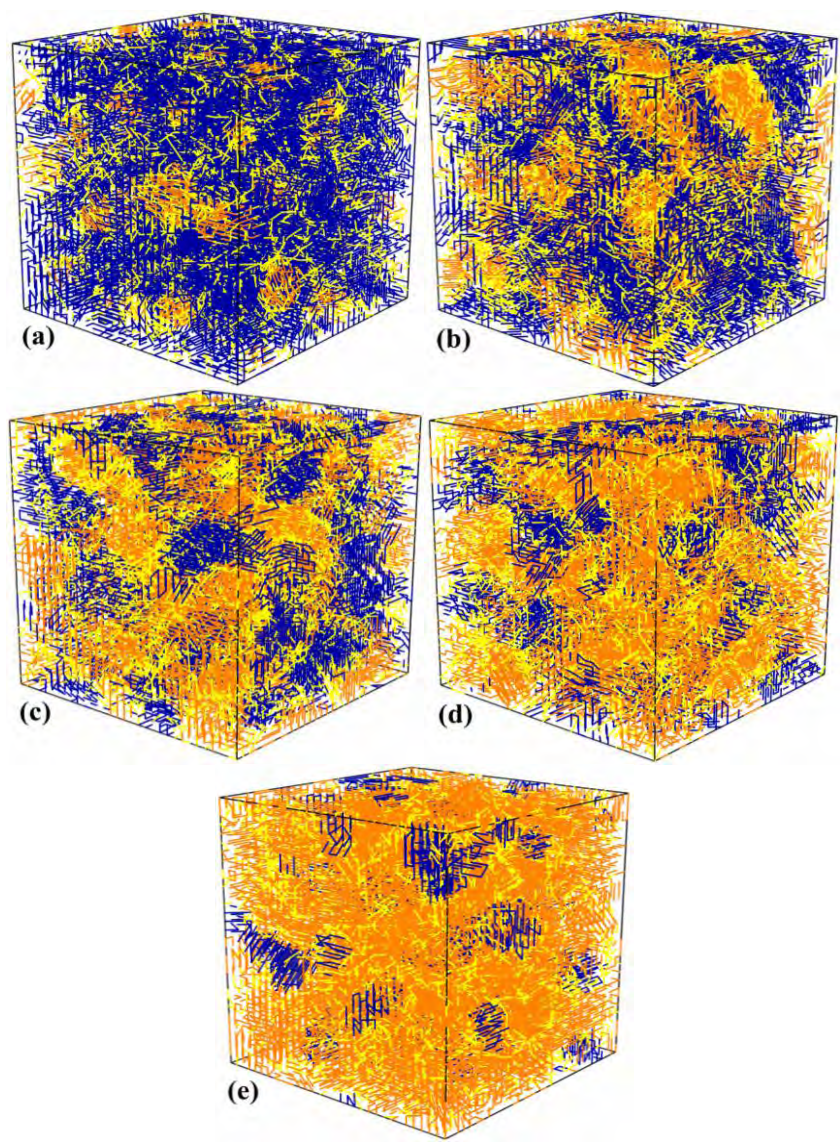

**Figure S15.** Snapshots of semi crystalline structure of diblock copolymer at $U_p$ = 0.6 for $x_B$ = **(a)** 0.125, **(b)** 0.375, **(c)** 0.50, **(d)** 0.625 and **(e)** 0.875 during one-step isothermal crystallization at $\lambda$ = 1. Blue and orange lines represent crystalline bonds of A-block and B-block respectively, and yellow lines represent non-crystalline bonds of both the blocks.